\documentclass[a4paper,11pt]{article}
\usepackage{jheppub} 

\usepackage[all,curve]{xy}
\input xy

\newtheorem{conjecture}{Conjecture}
\def\beq#1\eeq{\begin{align}#1\end{align}}

\title{Argyres-Douglas matter and  $\mathcal{N}=2$ dualities}

\author[b,c]{Dan Xie}
\author[a,b,c]{Shing-Tung Yau}

\affiliation[a]{Department of Mathematics, Harvard University, Cambridge, MA 02138, USA}
\affiliation[b]{Center of Mathematical Sciences and Applications, Harvard University, Cambridge, 02138, USA}
\affiliation[c]{Jefferson Physical Laboratory, Harvard University, Cambridge, MA 02138, USA}

\abstract{We study S duality of four dimensional $\mathcal{N}=2$ Argyres-Douglas (AD) theory  engineered from 6d $A_{N-1}$ $(2,0)$ theory. We find a  $(p,q)$ sequence of SCFTs, here $(p,q)$ is co-prime and class ${\cal S}$ theory defined on sphere corresponds to class $(0,1)$ theory. We represent these theories by a sphere with marked points, and S duality is interpreted as different pants decompositions of the same punctured sphere. The weakly coupled 
gauge theory description involves gauging AD matter which is represented by three punctured sphere. }

\begin{document} 
\maketitle
\flushbottom

\section{Introduction}
$S$ duality of four dimensional supersymmetric field theories has been extensively explored in the past several decades. 
The classical example is $\mathcal{N}=4$ supersymmetric Yang-Mills theory \cite{Montonen:1977sn}, where $S$ duality exchanges the gauge group $G$ 
and Langlands dual gauge group $G^{L}$. Similar $S$ duality has been found for $\mathcal{N}=2$ $SU(2)$ gauge theory coupled with four fundamental 
hypermultiplets \cite{Seiberg:1994aj}. For these examples, theories in different duality frames admit Lagrangian descriptions. 
Argyres-Seiberg \cite{Argyres:2007cn} generalized $S$ duality to $\mathcal{N}=2$ $SU(3)$ with six fundamental 
hypermultiplets, and the new feature is that one of the dual theory involves a strongly coupled matter system. 
Gaiotto \cite{Gaiotto:2009we} found a remarkable generalization  of Argyres-Seiberg duality by using 
6d $(2,0)$ construction. These so-called class  ${\cal S}$ theories tremendously improve the space of theories whose S duality behavior is known. 

One feature of class ${\cal S}$ theory is that the scaling dimension of Coulomb branch operator is integral. 
There are another class of $\mathcal{N}=2$ models called Argyres-Douglas (AD) theories whose Coulomb branch operators
have fractional scaling dimensions \cite{Argyres:1995jj,Argyres:1995xn}. These models seem to be much more general than the theory with integral scaling dimensions, 
and we would like to understand its S duality property. Two very interesting examples have been studied in \cite{Buican:2014hfa}, and 
several infinite class of self-dual models have been studied in \cite{DelZotto:2015rca}. Full S duality property of certain class of AD theories 
admitting 3d mirror has been found in \cite{Xie:2016uqq} using the decomposition of their 3d mirrors. 

The purpose of this paper is to find the S duality property of  all Argyres-Douglas  theories  engineered using 
6d $A_{N-1}$ type $(2,0)$  theories \cite{Xie:2012hs}. We first classify  AD theories with exact marginal deformations; and for AD theories without exact marginal deformations, we call them AD matter.  We found that it is natural to organize the theory space by a  $(p,q)$ label with $q>0, p\geq -q+1$, here $(p,q)$ is co-prime and class ${\cal S}$ theory defined on sphere  corresponds to class $(0,1)$ theory. 

\begin{center}
\begin{figure}[htbp]
\small
\centering
\includegraphics[width=3.5in]{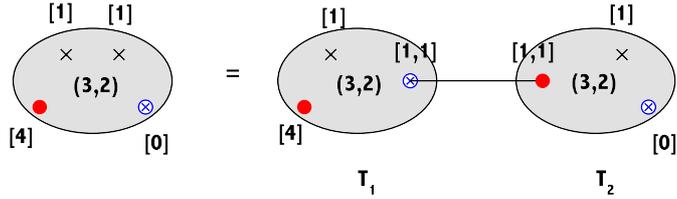}
\caption{$(A_3, A_5)$ theory is represented by a sphere with four  marked points, here we need to use three types of marked points besides the Young Tableaux data.
This theory belongs to  class $(3,2)$ theory, and its weakly coupled gauge theory description 
can be found from the degeneration limit of the punctured sphere. There are  two duality frames as we can only exchange two black marked points. }
\label{intro}
\end{figure}
\end{center}

$S$ duality of class ${\cal S}$ theory is elegantly solved by representing the SCFT by a Riemann surface with marked points \cite{Gaiotto:2009we}, and different $S$ duality 
frames correspond to different degeneration limits of punctured Riemann surface into three punctured sphere representing matter systems. Motived by  class ${\cal S}$ theory and the $S$ duality  of
AD theories found in \cite{Xie:2016uqq}, we successfully represent our class $(p,q)$ SCFT by a sphere with marked points: each punctured sphere has a label $(p,q)$, and each marked point is labeled by
a Young Tableaux $[a_1,a_2,\ldots, a_i]$ with arbitrary size (notice that for class S theory the size of Young Tableaux is the same for all the marked points), see left hand side of figure. \ref{intro} for an example.
Now we have  following  $S$ duality picture of our class $(p,q)$ theory: 
\begin{itemize}
\item The AD matter is represented by a sphere with three marked points.
\item  The number of exact marginal deformations of a SCFT is identified with the number of complex structure deformations of 
the punctured sphere.
\item  Different duality frames are represented by different degeneration limits of the same punctured sphere into three punctured sphere.
We can read off the field theory description of each duality frame from the degeneration limit. 
\end{itemize}
See figure. \ref{intro} for an example. We have made several checks such as the match of the Coulomb branch spectrum, central charges, and the vanishing of $\beta$ function, etc. 
There are several new features of  S duality of our general class $(p,q)$  theories: a): Generically (except class $(p,1)$ and $(1,q)$ theory) , there are three types of marked points (which we label them black, red and blue), and the AD matter is represented by  a sphere with one each of each color; This restricts the possible pants decompositions; b): Higher genus version is only possible for class ${\cal S}$ theory. 

This paper is organized as follows: section II  reviews basic facts of exact marginal deformations of 4d $\mathcal{N}=2$ SCFT; Section III 
describes the construction of $\mathcal{N}=2$ SCFT from  6d $A_{N-1}$ theory; Section IV describes the $S$ duality of  theories with 3d mirror, which is 
also called class $(p,1)$ theory; Section V describes $S$ duality of class $(p,q)$ theory; Section VI  discusses SCFT formed by conformally gauging AD matter, and we 
argue that higher genus version of general $(p,q)$ class theory is not possible. Finally a 
 conclusion is given in section VII.

\section{Exact marginal deformations of 4d $\mathcal{N}=2$ SCFT}
The representation theory of four dimensional $\mathcal{N}=2$ superconformal algebra (SCA) has been described in \cite{Dolan:2002zh}. $\mathcal{N}=2$ SCA  has an 
important $SU(2)_R\times U(1)_R$ R symmetry, and the half BPS operators have been classified in \cite{Dolan:2002zh}. There are two kinds of important half-BPS operators:
${\cal E}_{r,(0,0)}$ and $\hat{B}_R$, here $r$ is the $U(1)_R$ charge and $R$ is the integer specifying the $SU(2)_R$ representations. 
We are mostly interested in the operators ${\cal E}_{r,(0,0)}$ and its scaling dimension is given by the following formula
\begin{equation}
\Delta[{\cal E}_{r,(0,0)}]=r.
\end{equation}
If $r=2$, we have the following marginal deformation 
\begin{equation}
\delta S=\tau \int d^4x \tilde{Q}^4 {\cal E}_{2,(0,0)}+c.c.
\end{equation}
Such deformations are actually exact marginal, and it is proven in \cite{Argyres:2015ffa} that this type of deformations are 
the only exact marginal deformations for a $\mathcal{N}=2$ SCFT. 

Given a $\mathcal{N}=2$ SCFT, we might want to answer the following questions about exact marginal deformations:
\begin{itemize}
\item Counting the number of exact marginal deformations. 
\item Determine the Zamolodchikov metric on the conformal manifold \footnote{More precisely, the space of exact marginal deformation $\mathcal{N}=2$ SCFT is often not a manifold, but a moduli stack.}. 
\item Determine the weakly coupled gauge theory descriptions at  singularity of the conformal manifold.
\end{itemize}
The first question is easiest and can often be read from the Seiberg-Witten (SW) curve, while the second and third questions are significantly harder. The major task of this paper is to understand the 
first and the third question for $\mathcal{N}=2$ AD SCFT engineered from 6d $A_{N_1}$ $(2,0)$ theory.

\section{$\mathcal{N}=2$ SCFT from M5 branes}
A large class of four dimensional $\mathcal{N}=2$ SCFTs can be engineered by putting $6d$ $(2,0)$ theory of type $J$ on Riemann surface with various type of regular and irregular punctures. 
 A SCFT can be defined  using following configurations:
\begin{itemize}
\item A genus $g$ Riemann surface with arbitrary number of regular singularities.
\item A sphere with one irregular singularity.
\item A sphere with one irregular singularity and one regular singularity.
\end{itemize}
The classification of  SCFT is then  reduced to the classification of punctures. The regular singularity has been classified in \cite{Gaiotto:2009we,Nanopoulos:2009uw,Chacaltana:2012zy}, and is closed related to the classification of nilpotent orbit of Lie algebra $J$ \cite{collingwood1993nilpotent}. The irregular singularity has been classified in \cite{Xie:2012hs,Wang:2015mra}, and is related to the classification of positive grading of Lie algebra $J$ \cite{reeder2012gradings}.

\subsection{Classification of punctures }

\subsubsection{Irregular puncture}
Let's start with six dimensional  $A_{N-1}$ $(2,0)$ theory and compactify it on a Riemann surface $\Sigma$ to get 
a 4d $\mathcal{N}=2$ SCFT.  Hitchin's equation is defined on $\Sigma$ and the  classification of irregular singularity is reduced to the classification of  higher order singular  boundary condition of Hitchin's equation. 
Hitchin's equation involves a pair of fields $(A_\mu, \Phi)$ on $\Sigma$, and $\Phi$ is called Higgs field.  For irregular singularity, the Higgs field has the following behavior near the singularity \cite{Xie:2012hs,Wang:2015mra} (see also \cite{gaiotto2009wall,Cecotti:2011rv,Bonelli:2011aa}  for $A_1$ case):
\begin{equation}
\Phi={T_k\over z^{2+k/b}}+\sum_{-b\leq k^{'}<k} {T_{k^{'}}\over z^{2+k^{'}/b}};
\label{higgs}
\end{equation}
Here $z$ is the local coordinate near the singularity, and $T_k$ is a regular semi-simple element of Lie algebra $A_{N-1}$; $(b, k)$ is coprime, i.e. $(k,b)=1$.
We also ignore the regular terms which can freely fluctuate. The coefficients $T_i$ are semi-simple and can also be 
simultaneously diagonalized. In going around the puncture for one loop ( $z\rightarrow ze^{2\pi i}$), the Higgs field changes as follows
\begin{equation}
\Phi^{'}={exp^{-{2\pi i k \over b}}T_k\over z^{2+k/b}}+\sum_{-b\leq k^{'}<k} {exp^{-{2\pi i k^{'} \over b}}T_{k^{'}}\over z^{2+k^{'}/b}}.
\end{equation}
The above solution is consistent if one can find an inner automorphism $\sigma$ of Lie algebra $A_{N-1}$ such that 
\begin{equation}
\sigma T_{k^{'}} \sigma^{-1}=exp^{{2\pi i k^{'}  \over b}}T_{k^{'}}, 
\label{auto}
\end{equation}
So $\sigma \Phi^{'} \sigma^{-1}=\Phi$ up to the difference of regular terms. 
The above inner automorphism exists if the Lie algebra has the following grading \cite{reeder2012gradings}:
\begin{equation}
g=\bigoplus_{i\in Z/b} g^i.
\end{equation}
Let's take $k^{'}=n b+l$, with $0\leq l<b$, then $T_{k^{'}}\in g^l$.  

All such gradings are classified in \cite{elashvili2013cyclic}, which is 
related to the classification of \textbf{cyclic} element of the corresponding Lie algebra.
Here we give a brief review. Let's start with a nilpotent element $e$, then 
by the Morozov-Jacobson theorem, the element $e$ can be included in an $sl_2$-triple $(e,h,f)$, so that $[e,f] = h, [h,e] = 2e, [h,f] = -2f$. Then the eigenspace decomposition of $g$ with respect to ad $h$ action is a Z-grading of g:
\begin{equation}
g=\bigoplus_{j=-d}^d g_j.
\end{equation}
where $g_{\pm d}\neq 0$. The positive integer $d$ is called depth of the $Z$-grading. An element of g of the form $e+F$ , where $F$ is a non-zero element of $g_d$, is called a \textbf{cyclic} element, associated to $e$. 
A cyclic element is called \textbf{semi-simple} if the generic element of $e+F$ is semi-simple, and it is called \textbf{regular} semi-simple if $e+F$ is regular semi-simple.  

For each nilpotent orbit $e$, one can associate a weighted Dynkin diagram, and the weights take values in $0$, $1$ and $2$.  
A nilpotent element is called \textbf{even} if all the weights are  \textbf{even}. For each nilpotent orbit,  we can associate a positive integer
\begin{equation}
m={\sum a_i s_i+2}.
\end{equation}
Let $\epsilon$ be a primitive $m$th root of 1. For a nilpotent orbit, we can define an inner automorphism $\sigma_e$ of $g$ by letting 
\begin{equation}
\sigma_e(e_{\alpha_i})=\epsilon^{s_i} e_{\alpha_i},~~\sigma_e(e_{-\alpha_i})=\epsilon^{-s_i} e_{-\alpha_i},~i=1,\ldots,r.
\end{equation}
Where $e_{\pm \alpha_i}$ are Lie algebra elements  attached to roots $\pm \alpha_i$. This action is extended to the whole Lie algebra using the Cartan-Weyl basis. The automorphism $\sigma_e$ defines a $Z/m$ grading on Lie algebra
\begin{equation}
g=\bigoplus_{j\in Z/m} g^j;
\label{decom}
\end{equation}
Here $g^0=g_0$ is a reductive subalgebra of $g$. If $e$ is even, then the lowest non-zero part is
 $g^2=g_2+g_{-d}$, and all the odd part in the above decomposition is missing. So the order of automorphism is actually $m^{'}={m\over 2}$. 
 
Let's focus on $J=A_{N-1}$, and the Lie algebra is identified with the $N\times N$ traceless matrices. A nilpotent orbit is labeled by a Young Tableaux $[n_1,  \ldots, n_r]$ with $\sum n_i=N$, and  the corresponding $sl_2$ triple $(e,h,f)$ has the following standard form 
\begin{equation}
e=\left(\begin{array}{cccc}
J_{n_1}&0&0&0\\
0&J_{n_2} &0&0\\
0&0&\ldots &0 \\
0&0&0&J_{n_r} 
\end{array}\right), ~~~~
h=\left(\begin{array}{cccc}
D_{n_1}&0&0&0\\
0&D_{n_2} &0&0\\
0&0&\ldots &0 \\
0&0&0&D_{n_r} 
\end{array}\right). 
 \end{equation}
Here $J_{n_i}$ is the Jordan matrix with size $n_i$, and $D_{n_i}$ is the $n_i\times n_i$ dimensional diagonal matrix with diagonal entries $(n_i-1, n_i-3,\ldots,-(n_i-3),-(n_i-1)$.  The corresponding weighted Dynkin diagram is found as follows: rearrange the eigenvalues of $h$ such that 
they are monotonically decreasing $a_1\geq a_2 \geq  \ldots \geq a_{N-1}\geq a_N$, and the weighted Dynkin diagram is 
\begin{equation}
\overset{a_1-a_2}{\bullet}-\overset{a_2-a_3}{\bullet}\ldots- \bullet - \overset{a_N-a_{N-1}}{\bullet}
\end{equation}
The  semi-simple and regular semi-simple cyclic elements are classified in \cite{elashvili2013cyclic}, and the result is listed in table. \ref{cyclic}. 

\begin{table}
\begin{center}
\begin{tabular}{|c|c|c|}
  \hline
~ & Young Tableaux & order $m^{'}$ \\ \hline
      Semi-simple &$ [n_1,\ldots, n_1, 1,\ldots,1 ]$  & $n_1$ \\ \hline
        Regular semi-simple & $ [n_1,\ldots, n_1] $   &    $n_1$ \\ \hline
        ~&$[n_1,\ldots,n_1,1]$& $n_1$ \\ \hline
\end{tabular} 
\end{center}
\caption{The corresponding nilpotent orbit of semi-simple and regular semi-simple cyclic element of  Lie algebra $A_{N-1}$.}
\label{cyclic}
\end{table}

\textbf{Example 1}: Consider Lie algebra $sl_4$ and its  nilpotent orbit corresponding to partition $[4]$. The  corresponding $sl_2$ triple is 

\begin{equation}
e=\left(\begin{array}{cccc}
0&1&0&0\\
0&0 &1&0\\
0&0&0&1 \\
0&0&0&0
\end{array}\right), ~~
h=\left(\begin{array}{cccc}
3&0&0&0\\
0&1 &0&0\\
0&0&-1&0 \\
0&0&0&-3
\end{array}\right).
\end{equation}
So the weighted Dynkin diagram is $\overset{2}{\bullet}-\overset{2}{\bullet}-\overset{2}{\bullet}$.
After some computations, we find the following decomposition of lie algebra
\begin{align}
&g_{0}=\left(\begin{array}{cccc}
a_{11}&0&0&0\\
0&a_{22} & 0&0\\
0&0&a_{33}&9\\
0&0&0&a_{44}
\end{array}\right),
g_{2}=\left(\begin{array}{cccc}
0&a_{12}&0&0\\
0&0&a_{23}&0 \\
0&0&0&a_{34} \\
0&0&0&0 \\
\end{array}\right), 
g_{4}=\left(\begin{array}{cccc}
0&0&a_{13}&0\\
0&0 & 0&a_{24}\\
0&0&0&0 \\
0&0&0&0
\end{array}\right), 
g_{6}=\left(\begin{array}{cccc}
0&0&0&a_{14}\\
0&0 & 0&0\\
0&0&0&0 \\
0&0&0&0
\end{array}\right), \nonumber\\
&g_{-2}=\left(\begin{array}{cccc}
0&0&0&0\\
a_{21}&0&0&0 \\
0&a_{32}&0&0 \\
0&0&a_{43}&0 \\
\end{array}\right), 
g_{-4}=\left(\begin{array}{cccc}
0&0&0&0\\
0&0&0&0 \\
a_{31}&0&0&0 \\
0&a_{42}&0&0 \\
\end{array}\right), 
g_{-6}=\left(\begin{array}{cccc}
0&0&0&0\\
0&0&0&0 \\
0&0&0&0 \\
a_{41}&0&0&0 \\
\end{array}\right).
\end{align}
We have following grading on Lie algebra $sl(4)$:
\begin{equation}
g^{0}=g_0,~~g^2=g_{2}+g_{-6},~~g^4=g_{4}+g_{-4},~~g^6=g_{6}+g_{-2};
\end{equation}
and the order of actual automorphism is four, so we have a 
refined grading 
\begin{equation}
sl(4)=\bigoplus_{Z/4} g^l.
\end{equation}

\textbf{Example 2}: Consider $SL_4$ Lie algebra and the nilpotent element corresponding to the partition $[2,2]$; and the standard triple is 
\begin{equation}
e=\left(\begin{array}{cccc}
0&1&0&0\\
0&0 &0&0\\
0&0&0&1 \\
0&0&0&0
\end{array}\right), 
h=\left(\begin{array}{cccc}
1&0&0&0\\
0&-1 &0&0\\
0&0&1&0 \\
0&0&0&-1
\end{array}\right).
\end{equation}
Using the matrix $h$, we find that  the weighted Dynkin diagram is $\overset{0}{\bullet}-\overset{2}{\bullet}-\overset{0}{\bullet}$. Using $h$, we have the following decomposition of Lie algebra: 
\begin{equation}
g_{0}=\left(\begin{array}{cccc}
a_{11}&0&a_{13}&0\\
0&a_{22} & 0&a_{24}\\
a_{31}&0&a_{33}&0 \\
0&a_{42}&0&a_{44}
\end{array}\right),
g_{2}=\left(\begin{array}{cccc}
0&a_{12}&0&a_{14}\\
0&0&0&0 \\
0&a_{32}&0&a_{34} \\
0&0&0&0 \\
\end{array}\right),
g_{-2}=\left(\begin{array}{cccc}
0&0&0&0\\
a_{21}&0 & a_{23}&0\\
0&0&0&0 \\
a_{41}&0&a_{43}&0
\end{array}\right).
\end{equation}
The Lie algebra has the following grading
\begin{equation}
g^0=g_0,~~g^2=g_2+g_{-2}.
\end{equation}
and the order of actual automorphism is two:
\begin{equation}
sl_4=\bigoplus_{Z/2} g^l.
\end{equation}

\textbf{Classifications}: Now let's explain how to define our irregular singularity using the grading of Lie algebra. 
Start with $J=A_{N-1}$, and given a Young Tableaux $[n_1,\ldots, n_1]$ or $[n_1, \ldots, n_1,1]$ which define a regular semi-simple cyclic element,   the Higgs field takes the form:
\begin{equation}
\Phi={T_k\over z^{2+{k\over n_1}}}+\sum_{-{n_1}\leq k^{'}<k} {T_{k^{'}}\over z^{2+{k^{'}\over n_1}}};
\end{equation}
Here $T_{k^{'}} \in g^l$ with $k^{'}=n n_1+l$ and $ 0 \leq l< n_1$.  
They are all taken to be diagonal matrices, and the eigenvalues take the following form
\begin{align}
& T_i=\text{diag}(a_1 A_i,\ldots, a_2 A_i,\ldots, a_{{N\over n_1}} A_i)~~~~~~Y=[n_1,\ldots, n_1] ,\nonumber\\
&T_i=\text{diag}(a_0, a_1 A_i,\ldots, a_2 A_i,\ldots, a_{{N\over n_1}} A_i)~~~~Y=[n_1,\ldots, n_1,1].
\label{degen}
\end{align}
Here $A_i$ are fixed diagonal  matrix with size $n_1$, and traceless condition would force some coefficients $a_i$ to be zero.  
The detailed classification of SCFT is given as follows:
\begin{enumerate}
\item Let's take the trivial grading corresponding to partition {$[1,\ldots, 1]$},  
then the corresponding irregular singularity is 
\begin{equation}
\Phi={T_n\over z^n}+\ldots +{T_{n-1}\over z^{n-1}}+\ldots+{T_{1}\over z}.
\label{trivial}
\end{equation} 
Here $T_i$s are diagonal matrix with distinct eigenvalues.
\item Let's take the grading corresponding to partition $[N]$, the Higgs field takes the following form
\begin{equation}
\Phi={T\over z^{2+k/N}}+\ldots
\end{equation}
Here  $k$ and $N$ are coprime.  More generally, we can choose the grading corresponding to the partition $[n_1,\ldots, n_1]$ with $n_1$ a divisor of $N$, and the Higgs field takes the form
\begin{equation}
\Phi={T\over z^{2+k/n_1}}+\ldots
\end{equation}
One can also engineer this class of theory using 
type IIB string theory on a 3-fold singularity of the following form \cite{Cecotti:2010fi, Xie:2012hs}
\begin{equation}
x_1^2+x_2^2+x_3^N+z^{{N\over n_1}k}=0.
\end{equation}

\item  Let's take the grading corresponding to partition $[N-1,1]$, the Higgs field  takes the form 
\begin{equation}
\Phi={T\over z^{2+{k\over N-1}}}+\ldots
\end{equation}
here $k$ and $N-1$ is coprime. More generally, one can take the grading corresponding to  partition $[n_1,\ldots, n_1, 1]$ with $n_1$ a divisor of $N-1$, and the Higgs field takes the following form
\begin{equation}
\Phi={T\over z^{2+{k\over n_1}}}+\ldots
\end{equation}

One can also engineer this class of theory using 
type IIB string theory on a 3-fold singularity of the following form \cite{Xie:2012hs}:
\begin{equation}
x_1^2+x_2^2+x_3^N+x_3z^{{N-1\over n_1}k}=0.
\end{equation}
\end{enumerate}

\textbf{Remark}: It is interesting to explore whether one could get interesting SCFT using general semi-simple grading listed in table. \ref{cyclic}.

\textbf{Degenerations}: 
Let's first consider the irregular singularity corresponding to the partition $[1,\ldots, 1]$, see (\ref{trivial}).
Here $T_i$ are taken to be diagonal matrices. The degeneration is defined by a sequence of Levi sub-algebra $L_1\subset L_2 \subset \ldots \subset L_n$ so that $L_i$ commutes 
with the set $(T_n,T_{n-1},\ldots, T_{n-i})$ \cite{Witten:2007td}.  Let's use a sequence of Young Tableaux $Y_i$ to denote the degeneracy of 
the eigenvalue of $T_n$, then the constraints coming from the inclusion relation of the Levi sub-algebra is that the Young Tableaux $Y_i$ is 
derived by decomposing the columns of $Y_{i+1}$. 

One can generalize the above consideration to the situation corresponding to  partition $[n_1,\ldots, n_1]$ or $[n_1, \ldots, n_1, 1]$. For the partition $[n_1,\ldots, n_1]$, the irregular singularity is 
\begin{equation}
\Phi={T_k\over z^{2+{k\over n_1}}}+\sum_{-{n_1}\leq k^{'}<k} {T_{k^{'}}\over z^{2+{k^{'}\over n_1}}};
\end{equation}
Each diagonal matrix $T_i$ takes the form in (\ref{degen}), and one can use a Young Tableaux with size ${N\over n_1}$ to denote the eigenvalue degeneracy of the coefficient $a_i$. Now 
the degeneracy can be defined by choosing a sequence of  $k+n_1+1$ Young Tableaux with size ${N\over n_1}$:
\begin{equation}
Y_{k}\subset Y_{k-1} \ldots \subset Y_{-n_{1}}.
\end{equation} 
Similarly, one have a sequence of Young Tableaux to describe the degeneration of irregular singularity defined using partition $[n_1,\ldots, n_1, 1]$. 
Note however not all the irregular singularity defined from degeneration define a SCFT, and those which actually define SCFTs will 
be classified later.

\subsubsection{Regular puncture}
One still get AD SCFT if we add an extra regular singularity besides an irregular singularity. 
The regular singularity for $A_{N-1}$ Hitchin system has been classified in \cite{Gaiotto:2009we}, and the classification coincides with 
the classification of nilpotent orbit which  is labeled by a Young Tableaux  $[n_s^{h_s},\ldots, n_1^{h_1} ]$ with $n_s>\ldots>n_1$, whose
 flavor symmetry is 
\begin{equation}
G_{Y}=[\prod_{i=1}^s U(h_i)]/U(1).
\end{equation}
The favor central charge depends on the detailed form of the other irregular singularity and will be computed later. 

\subsection{Newton Polygon and SW curve}
\begin{center}
\begin{figure}[htbp]
\small
\centering
\includegraphics[width=6cm]{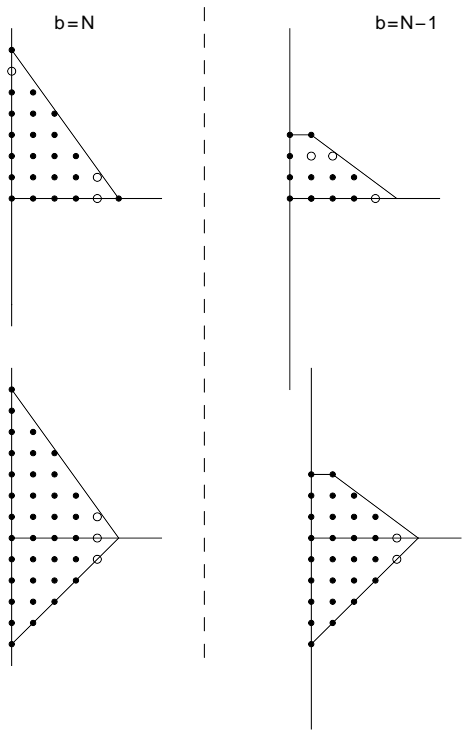}
\caption{Newton polygon of SCFT defined by 6d $A_{N-1}$ $(2,0)$ theory on a sphere with an irregular singularity and a regular singularity. The SW curve can be found from the monomials associated with black bullets within Newton polygon.}
\label{torus}
\end{figure}
\end{center}

One can represent the irregular singularity (\ref{higgs}) of $A_{N-1}$ theory by a Newton polygon, see figure. \ref{torus}. The slop of the boundary line 
represents the leading order of pole (minus two) of the irregular singularity. 
A regular singularity adds a further boundary line which connects point $(-N,0)$ and point $(0,N)$.  
The SW curve is identified with the spectral curve of the Hitchin system 
\begin{equation}
det(x-\Phi)=0.
\end{equation}
The SW curve of Coulomb branch can be easily found from Newton polygon, i.e. we associate to a point with coordinate $(m,n)$ a monomial $x^m z^n$, and the SW curve  is simply (Consider theory defined using partition $[n_1,\ldots, n_1]$)
\begin{equation}
x^N+z^k+\sum_{(m,n)\in S} u_{m,n} x^m z^n=0,
\end{equation}
where the coefficients $u_{m,n}$ label the parameters (couplings, Coulomb branch operators and masses) of the AD theory. Notice that we only count square points within the Newton polygon. 
One can find the scaling dimensions of these parameters by demanding each term in SW curve to have the same scaling dimension and  that the SW differential $\lambda=x dz$ has the scaling dimension one. 
Consider theory defined using partition $[n_1,\ldots, n_1]$, the SW curve of the AD point is 
\begin{equation}
x^k+z^N=0.
\end{equation}
Then we have the following two equations
\begin{equation}
k[x]=N[z],~~[x]+[z]=1.
\end{equation}
and it is easy to find  $[x]={N\over N+k},~~[z]={k\over N+k}$, then it is straightforward to find the full spectrum of the theory.

\subsection{Argyres-Douglas matter}
One can read off the number of exact marginal deformations and the mass parameters from the SW curve. On the other hand, one can read those numbers from 
the data in defining irregular singularity: the number of parameters in leading order matrix $T_k$ gives the maximal possible exact marginal deformations, and  the number of mass parameters 
are identified with the parameters in diagonal matrix $T_0$. For the AD theories using the regular semi-simple cyclic element, those numbers  are listed in table. \ref{mass}. 
\begin{table}[!htb]
\begin{center}
\begin{tabular}{|c|c|c|}
  \hline
 Young Tableaux & exact marginal parameters & mass parameters  \\ \hline
    $ [n_1,\ldots, n_1]$ & ${{N\over n_1}}-1$ &  ${{N\over n_1}}-1$ \\ \hline
       $ [n_1,\ldots, n_1,1]$ & ${{N\over n_1}}-1$ & ${{N\over n_1}}-1$ \\ \hline
\end{tabular} 
 \caption{Number of maximal possible exact marginal deformations and mass parameters for the AD theories defined by the corresponding cyclic element of $A_{N-1}$ Lie algebra. }
       \label{mass}
\end{center}
\end{table}

Argyes-Douglas matters are defined as  those SCFTs satisfying the following two questions: a) there are no exact marginal deformations. 
b) there is non-abelian flavor symmetry; A detailed classification will be given in following sections.

\section{Theory of class $(p,1)$} 
Let's consider 4d $\mathcal{N}=2$ SCFTs defined by compactifying 6d $(2,0)$ $A_{N-1}$ theory on a sphere with following irregular singularity: 
\begin{equation}
\Phi={T_n\over z^n}+\ldots+{T_1\over z}.
\end{equation}
The eigenvalue degeneracies of  matrices $T_i$ are encoded by a sequence of Young Tableaux
\begin{equation}
Y_n\subset Y_{n-1} \subset \ldots \subset Y_1.
\label{irrep1}
\end{equation}
This class of theories are defined using the partition $[1,\ldots, 1]$, see the notation in last section. 
We also consider theories defined using above irregular singularity plus an extra regular singularity, which is denoted by 
a Young Tableaux $Y_0$. We  call these models as class $(p,1)$ theories with $p=n-2$, and the reasoning will be clear in next section. 
A theory might be denoted by $(A_{N-1}; Y_{n},\ldots, Y_1; Y_0)$; Not all of them describe a SCFT and we will 
classify those configurations which actually define SCFTs. 

A special feature of these theories is that they admit 3d mirror with Lagrangian descriptions. The 3d mirror is quite useful in
understanding many properties of  original 4d theory. In particular, it was shown in \cite{Xie:2016uqq} that one can interpret $S$ duality of 4d theory as different 
decompositions of 3d mirror. In \cite{Xie:2016uqq}, we outline the main idea of finding $S$ duality  for this class of theories. 
Here we will introduce further combinatorial tools and provide a systematical study of $S$ duality of above SCFTs.

\subsection{Coulomb branch spectrum}
Let's first review how to find  Coulomb branch spectrum for this class of theories \cite{Xie:2012hs}. 
The SW curve is identified with the spectral curve of the corresponding HitchinÔs system:
\begin{equation}
\text{det}(x-\Phi)=0\rightarrow x^N+\sum_{i=2}^N \phi_i (z)x^{N-i}=0.
\end{equation}
Here $z$ denotes the coordinate on Riemann surface $\Sigma$ on which we compactify 6d $(2,0)$ theory. 
At the SCFT point, the SW curve takes the following form,
\begin{equation}
x^N+z^{(n-2)N}=0,
\label{sw1}
\end{equation}
from which we can find out the scaling dimension of $x$ and $z$:
\begin{equation}
[x]={n-2\over n-1},~~~[z]={1\over n-1}.
\end{equation}
This is determined by requiring the SW differential $\lambda=xdz$ having scaling dimension one, and requiring each term in 
(\ref{sw1}) to have the same scaling dimension. 

\begin{figure}[h]
\centering
  \includegraphics[width=4in]{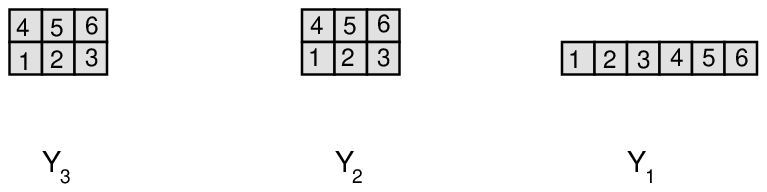}
  \caption{Young Tableaux with labels. Here $Y_3=[2,2,2],Y_2=[2,2,2], Y_1=[1,1,1,1,1,1]$.}
  \label{young}
\end{figure}

The Coulomb branch spectrum can be computed as follows: Let's consider one of Young Tableaux $Y_j= [n_1, \ldots, n_{r_{j}}]$ in the definition of irregular singularity (\ref{irrep1}), and label the boxes 
of $Y_j$ as $1,\ldots N$ starting from the bottom-left corner of $Y_j$ and going row by row, see figure. \ref{young}; Define a sequence of  integral numbers $p_i^{(j)}$:
\begin{equation}
p_i^{(j)}=i-s_i^{(j)},~~i=1,\ldots, N
\label{seqnum}
\end{equation}
here $s_i^{(j)}$ is the height of the $i$th box in $Y_j$. The independent Coulomb branch operators in $\phi_i(z)$ are found from coefficients of the monomials $ z^{k}x^{N-i}, 0\leq k\leq d_i$, and 
$d_i$ is given by the following formula
\begin{equation}
d_i=\sum_{j=1}^{n}p_i^{(j)}-2i.
\label{spec}
\end{equation}

\textbf{Example}: Consider an irregular singularity specified by the Young Tableaux $Y_3=[2,2,2], Y_2=[2,2,2],~Y_1=[1,1,1,1,1,1]$.  Using formula \ref{seqnum}, we get following set of numbers:
\begin{equation}
p_i^{(3)}=(0,1,2,2,3,4),~~p_i^{(2)}=(0,1,2,2,3,4),~~p_i^{(1)}=(0,1,2,3,4,5).
\end{equation}
So the set of numbers $d_i$ are 
\begin{equation}
d_i=(-2,-1,0,-1,0,1).
\end{equation}
Negative number means that there is no coulomb branch deformation in the corresponding differential $\phi_i$. So the SW curve with independent Coulomb branch operators are 
\begin{equation}
x^6+ c_1 x^3+ c_2 x+ (z^6+c_3 z+c_4)=0. 
\end{equation}
The scaling dimensions of these Coulomb branch operators are:
\begin{equation}
[c_1]={3\over 2},~~[c_2]={5\over 2},~~[c_3]={5\over2},~~[c_4]=3.
\end{equation}

Now let's add an extra regular singularity which is labeled by a Young Tableaux $Y^0=[n_1,\ldots, n_{r_0}]$, and again define a sequence of numbers  $p_0$
using formula \ref{seqnum}. The maximal order of pole in $\phi_i$ is just $p_i^{(0)}$.  The independent Coulomb branch operators in $\Phi_i(z)$ are found from the 
coefficients before following monomials
\begin{equation}
 z^k x^{N-i},~~~~~-p_i^{(0)}\leq k\leq d_i. 
\end{equation}

\subsection{3d Mirror}
Let's compactify a  four dimensional $\mathcal{N}=2$ theory on a circle and flow to IR to get a 3d $\mathcal{N}=4$ SCFT $A$. For 
a 3d $\mathcal{N}=4$ SCFT $A$, one can often find a mirror SCFT $B$ \cite{Intriligator:1996ex}. The basic feature of the 3d mirror is that the Coulomb branch of 
theory $A$ is mapped to Higgs branch of theory $B$, and vice versa.    

\begin{figure}[htbp]
\small
\centering
\includegraphics[width=12cm]{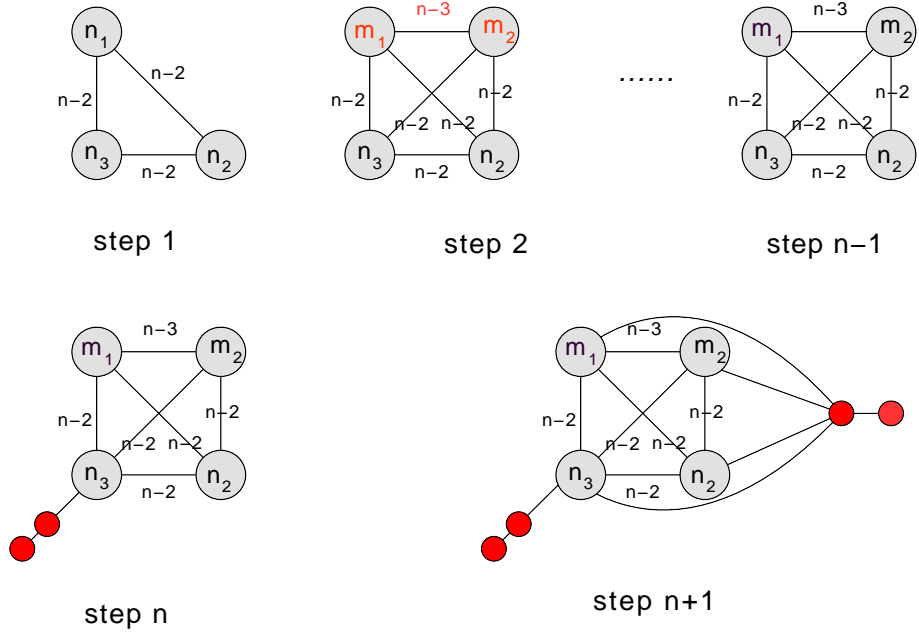} 
\caption{Step 1: If the first Young Tableaux $Y_n$ has partition $[n_1, n_2, n_3]$, first assign a quiver  with three nodes and ranks $n_i$, and then connect $n-2$  quiver arrows
between those nodes. Step 2: If  $n_1$ is further partitioned into $[m_1, m_2]$ in $Y_{n-1}$, we split the quiver node with rank $n_1$ into two quiver nodes 
with rank $m_1$ and $m_2$, and the number of quiver arrows between $m_i$ and $n_1$, $n_2$ are still $n-2$; but  the number of arrows between $m_1$ and $m_2$ are $n-3$. 
Similar procedure is done for other Young Tableaux and we stop at $Y_2$. Bottom: If a  column with height $l$ in $Y_2$ is further split into $[l_1, l_2,\ldots,l_t]$, one attach 
a quiver tail to the node with rank $l$  in quiver  determined by $(Y_n,\ldots Y_2)$; For $Y_0$, we attach a quiver tail which is connected to all the quiver nodes determined by $Y_2$.
The number of quiver arrow is one if there is no label.  }
\label{3dge}
\end{figure}

The three dimensional mirror theory $B$ of our model is found in \cite{Xie:2012hs}, and they all admit a Lagrangian description. 
 Consider a theory denoted as $(A_{N-1}; Y_{n},\ldots, Y_1; Y_0)$, and  its 3d mirror is derived step by step as the following:
 \begin{enumerate}
\item  Assume $Y_n =[n_1,n_2,....n_r]$, then assign $r$ quiver nodes with
gauge group $U(n_i)$, and draw $n-2$ quiver arrows connecting any pair of quiver nodes, see figure. \ref{3dge}.

\item 
If one of column with height $n_i$ of $Y_{n}$ is further partitioned as $[m_{i1}, m_{i2},\ldots m_{is}]$ in Young  Tableaux $Y_{n-1}$, we split the quiver node with rank $n_i$ into  several quiver nodes with rank $m_{ij}$, and draw $n-3$ 
quiver arrows between those new created quiver nodes,  see figure. \ref{3dge} step 2.
One do the similar splitting for each Young Tableaux until  $Y_2$, and  get a quiver  with many quiver nodes and nested arrows 
between them. Notice that the sum of the total rank  of all the quiver nodes are $N$.

\item
The special treatment is needed for $Y_1$:  if  one of the column of $Y_2$ has height $l$ and 
is further partitioned as $[l_1,l_2,\ldots ,l_t]$ in $Y_1$,  we do not split the quiver node with rank $l$, 
instead we attach a quiver tail as follows: define $h_i=\sum_{t}^{j=t-i+1} l_j$, the quiver tail is
\begin{equation}
U(h_1)-U(h_2)-\ldots-U(h_{t-1})-[U(l)].
\end{equation}

\item If there is an extra regular singularity specified by a Young Tableaux $Y_0=[p_1,p_2,\ldots, p_q]$, define $h_i=\sum_{q}^{j=q-i+1} p_j$, and assign a quiver tail:
\begin{equation}
U(h_1)-U(h_2)-\ldots-U(h_{q-1})-[U(N)].
\end{equation}
Then spray the $U(N)$ node as the pattern determined by the Young Tableaux $Y_2$ of the  the irregular singularity. 
Finally we glue the quiver of irregular singularity and regular singularity by identifying the sprayed nodes of regular singularity tail with the quiver nodes determined by $Y_2$.

\end{enumerate}

\begin{figure}[h]
\centering
  \includegraphics[width=4in]{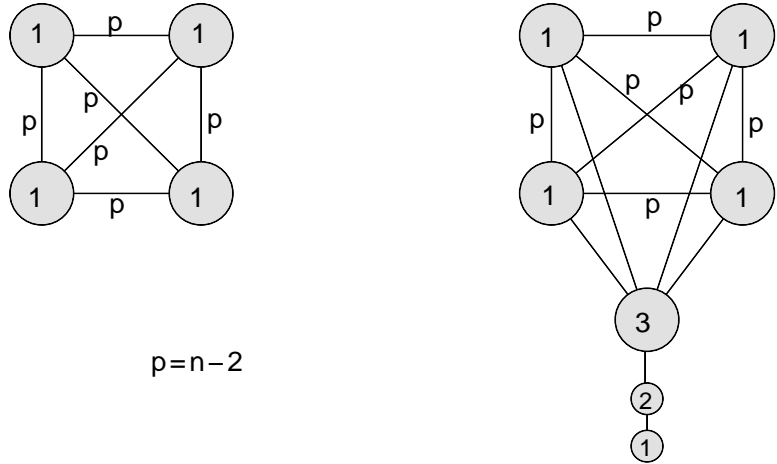}
  \caption{Left:  3d mirror for theory $(A_{3}; Y_{n}, \ldots, Y_{1}; Y_0)$ with $Y_n=\ldots=Y_1=[1,1,1,1]$ and $Y_0=[4]$ (trivial). Right: 3d mirror for theory $(A_{3}; Y_{n}, \ldots, Y_{1}; Y_0)$ with $Y_n=\ldots=Y_1=[1,1,1, 1]$ and $Y_0=[1,1,1,1]$.
  The number of quiver arrow is one if there is no label.}
  \label{3d1}
\end{figure}

Two examples are shown in figure. \ref{3d1}. The 3d mirror can be used to learn interesting physical properties of original 4d theory:
\begin{itemize}
\item The 3d mirror can be used to find the non-abelian flavor symmetry of the original 4d theory. The Higgs branch of our 4d theory is mapped to the Coulomb branch of 
the 3d mirror theory $B$, and the flavor symmetry of 4d theory can be read from the Coulomb branch symmetry of theory $B$. The method of reading flavor symmetry of Coulomb branch of a quiver gauge theory 
is developed in \cite{Gaiotto:2008ak}. Here we review the basic ingredients. A quiver node is defined as \textbf{balanced} if $N_f=2N_c$. For each balanced ADE chain (which can only be ADE shape), we have 
a corresponding ADE flavor group on Coulomb branch, and there is a $U(1)$ flavor symmetry for each non-balanced quiver node. 
\item  Sometimes the original 4d theory is not irreducible, i.e. the 4d theory consists an interacting part and a free part. One can use 3d mirror to detect it and actually find the interacting part \cite{Nanopoulos:2010bv}. Let's 
define a quiver node to be \textbf{bad} if $N_f<2N_c$, and our original 4d theory is reducible if the mirror quiver has a bad node. To get the interacting part of original 4d theory, we perform the following 
operation: first replace the rank  of a bad node as follows $N_c^{'}=N_f-2N_c$. After this step one may create other bad nodes, and one need to continue performing the above operation until there 
is no bad node left. The final quiver describes the mirror of the interacting part of original theory. 
\item Given a 3d mirror quiver $B$, one can reverse engineer the M5 configuration of theory $A$. Often theory A can have more than one M5 brane constructions \cite{Xie:2012hs, boalch2008irregular}. 
\end{itemize}

\subsection{Classification of SCFT}
We would like to identify theory $(A_{N-1}; Y_n,\ldots, Y_1; Y_0)$ which defines a SCFT. One necessary condition is that the possible number 
of exact marginal deformations should be larger than the number of dimension two operators. It appears that this is also a sufficient condition for our model.

Let's first assume that  $n>3$ and $Y_n$ has $a$ columns. Then the maximal 
number of exact marginal deformations are $a-2$: there are $a$ parameters in defining $Y_n$, and traceless condition and an overall scaling removes two parameters. 
Let's now count the number of dimension two operators. The SW curve is 
\begin{equation}
x^N+\sum_{i=2}^N \phi_i(z) x^{N-i}=0.
\end{equation}
The dimension two operator appears as the coefficient of the following monomial in SW curve:
\begin{equation}
z^{b_i}x^{N-i},~~~b_i=(n-2)i+2-2n.
\end{equation}
Here we need $i\geq 3$. Let's  look at differential $\phi_i, 3\leq i\leq  a$, and the maximal order of $z$ in $\phi_i$ which will give  Coulomb branch operator  is 
\begin{equation}
d_i=n(i-1)-2i=(n-2)i-n,~~~~i\leq a.
\end{equation} 
We have used formula (\ref{spec}) and the fact that all the Young Tableaux  has at least $a$ columns, so each Young Tableaux contributes $i-1$ to $d_i$.  We have $d_i-b_i=n+2>0$, and this  implies that there are 
at least $a-2$ dimensional two operators in the Coulomb branch spectrum.

Next let's consider differential $\phi_{a+1}$, and the maximal order of $z$ which will give Coulomb branch operator is 
\begin{equation}
d_{a+1}= n_1(a+1-2)+n_2(a+1-1)-2(a+1)=(n-2)(a+1)-2n+n_2.
\end{equation}
Here $n_1$ (resp. $n_2$) is the number of Young Tableaux which has $a$ (resp. more than $a$) columns, and $n_1+n_2=n$.    
To require $d_{a+1}<b_{a+1}$, we get $n_2<2$, so we only have zero or one Young Tableaux with more than $a$ columns. This implies that the first $n-1$ Young Tableaux 
all has $a$ columns, which means that they are all equal (Notice that $Y_i$ is derived by further splitting the columns of $Y_{i+1}$, so if $Y_{i+1}$  and $Y_i$ have the same columns, they are equal.).
The above result is not changed by adding an extra regular singularity.  Moreover, once we have $Y_n=\ldots=Y_2$, there is no dimension two operator in differential $\phi_i, i>a$. So 
the number of dimension two Coulomb branch operator matches with the number of exact marginal deformations. 

\begin{figure}[h]
\centering
  \includegraphics[width=6in]{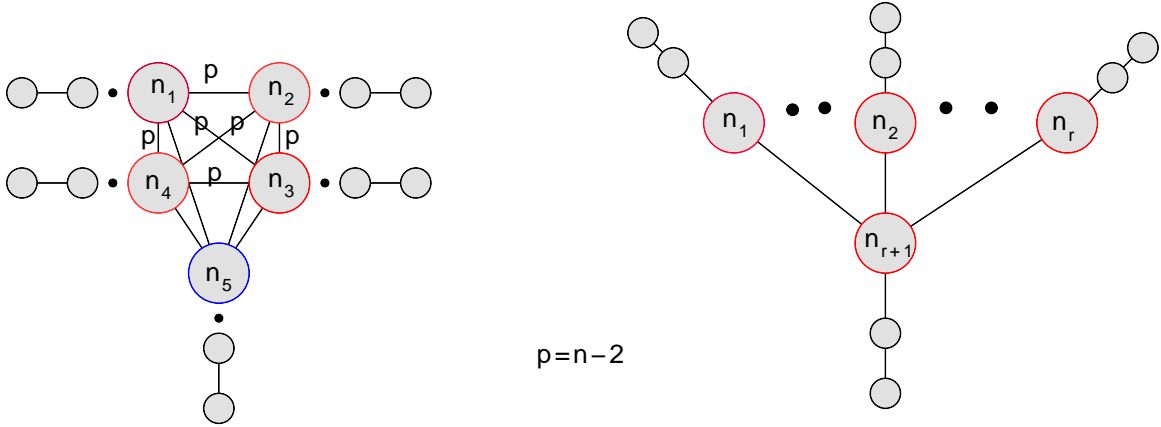}
  \caption{The 3d Mirror of class $(p,1)$ SCFT. These quiver nodes with explicit ranks form the core of the quiver, and those nodes without rank form a quiver tail. 
   Left:~$Y_n=\ldots=Y_2=[n_1,n_2, n_3, n_4]$ with $n\geq 3$,  and $Y_1,Y_0$ are arbitrary. Right:~$n=2$, and $(Y_2,~Y_1,~Y_0)$ are arbitrary. }
  \label{scft}
\end{figure}

Now let's consider the situation $n\leq 3$, after careful analysis, we found the following two possibilities 
\begin{itemize}
\item $Y_3$ and $Y_2$ both have $a$ columns, and $Y_1$ is arbitrary, the 3d mirror is shown on the left of figure. \ref{scft}, and $p=1$.
\item   $Y_2$ and $Y_1$ both have arbitrary Columns, and $Y_0$ is arbitrary \footnote{Such theories have multiple realizations using different six dimensional theory and irregular singularities, here we choose 
a realization with maximal number of possible exact marginal deformations.}. The 3d mirror is a star shape quiver, see figure. \ref{scft}. 
\end{itemize}
For above class of theories, there are only $a-3$ dimension two operators. 
The 3d mirror for these SCFTs is shown in figure. \ref{scft}. The 3d mirror on the left of figure. \ref{scft} is actually 3d mirror of class ${\cal S}$ theory defined on sphere. 

Among these SCFTs, the 3d mirror for AD matter  is shown in figure. \ref{poneAD}.
Let's make some comments about these AD matters:
\begin{figure}[h]
\centering
  \includegraphics[width=5in]{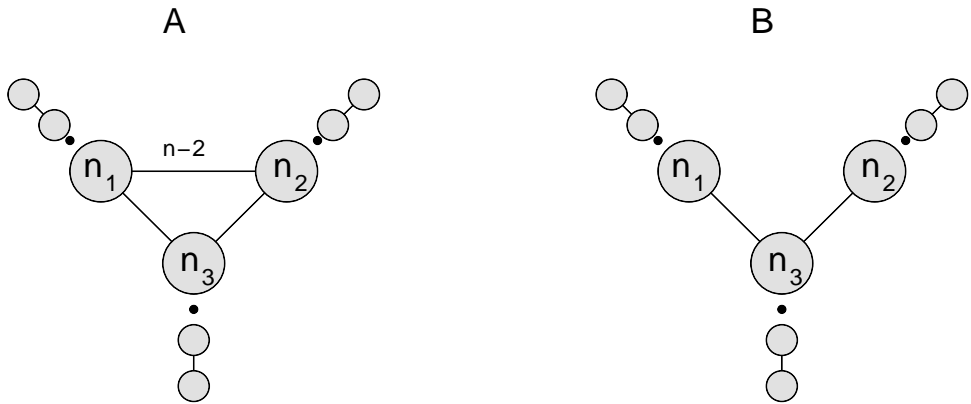}
  \caption{The 3d mirror of AD matter in class $(p,1)$ theories, here $p=n-2$.}
  \label{poneAD}
\end{figure}

\begin{figure}[h]
\centering
  \includegraphics[width=4in]{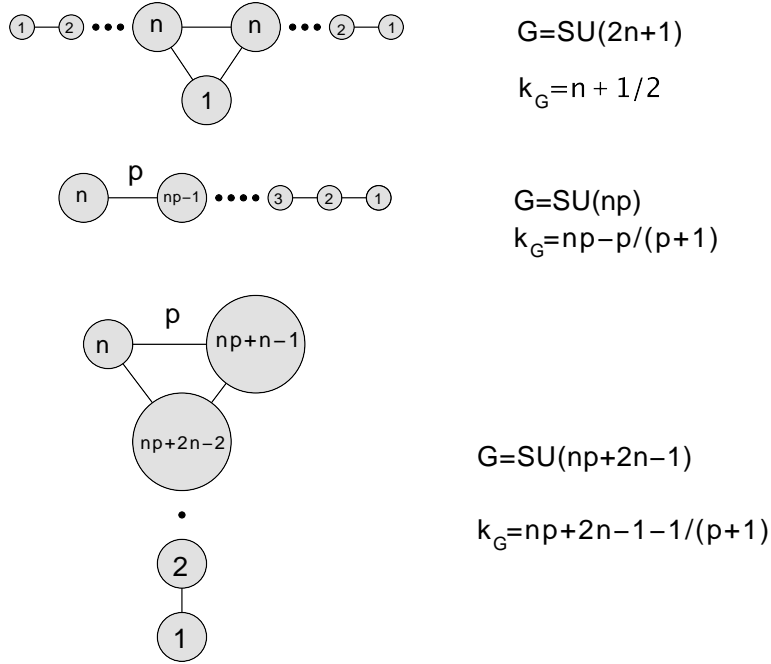}
  \caption{3d mirror of AD matter whose flavor symmetry is a simple $SU$ group, and we also list the flavor central charge.}
  \label{enhanced}
\end{figure}

\begin{enumerate}
\item Let's choose numbers $(n_1, n_2, n_3)$ in figure. \ref{poneAD} so that the 3d mirror is a good quiver, then the maximal flavor symmetry of AD matter on the left can be $SU(n_1)\times SU(n_2)\times SU(n_3)$ (further enhancement is also possible). The flavor central charge for them is 
\begin{equation}
k_{SU(n_1)}=n_1+{1\over n-1},~~~k_{SU(n_2)}=n_2+{1\over n-1},~~~~k_{SU(n_3)}=n_3+{n-2\over n-1}.
\end{equation}
The flavor symmetry $SU(n_3)$ can be further enhanced to $SU(n_3+1)$ with the same flavor central charge. 

For matter system on the right of figure. \ref{poneAD}, to have a good quiver, we need $n_3>n_1$ and $n_3>n_2$. Let's fix $n_3$, the maximal flavor symmetry can be $SU(n_3)\times SU(n_3)\times SU(n_3)$. This is actually
$T_{n_3}$ theory, and the flavor central charge is 
\begin{equation}
k_{SU(n_3)}=n_3.
\end{equation}
We can consider other type of flavor symmetries which is the subgroup of $SU(n_3)\times SU(n_3)\times SU(n_3)$. 

\item The AD matter could have enhanced flavor symmetry. The AD matter whose  flavor symmetry is enhanced to a single $SU$ group is shown in figure. \ref{enhanced}. 

\item  The Coulomb branch spectrum can be computed using the defining data of the irregular singularity and regular singularity, see formula (\ref{spec}). 
The central charge $a$ and $c$ can be computed using the following formula
\begin{equation}
2a-c={1\over 4}\sum_{i=1}^r(2 u_i-1),~~a-c=-{d_H\over 24}.
\label{central}
\end{equation}
Here $r$ is the dimension of Coulomb branch, and $d_H$ is the Higgs branch dimension which is equal to the dimension of Coulomb branch of 3d mirror.

\end{enumerate}

\begin{figure}[h]
\centering
  \includegraphics[width=2.5in]{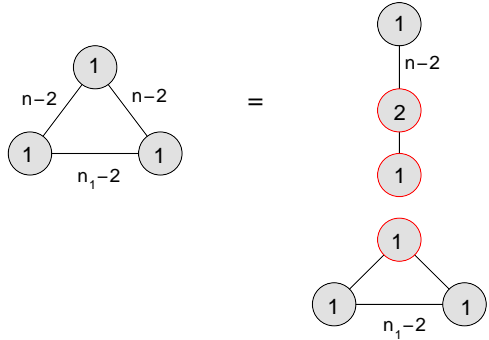}
  \caption{3d mirror of non-conformal theory, here $n_1>1$. The field theory description can be found from decomposing the 3d mirror into two pieces, an the field theory
 description  is $T_1-SU(2)-T_2$. Here $T_1$ (resp. $T_2$ ) matter contribute $2+{1\over n-2}$ (resp. $1+{n_1-1\over n_{1}-2}$ or 2 for $n_1=2$. ) to the $\beta$ function of $SU(2)$ gauge group, so the total 
 contribution of matter to the $\beta$ function is less than four, and the theory is an asymptotical free theory.}
  \label{nonconformal}
\end{figure}

\textbf{Remark}:  An interesting question is the interpretation of other non-conformal theories: they admit four dimensional $\mathcal{N}=2$ SUSY but not the full superconformal 
symmetry; some of them can be described by asymptotical free gauge theories, while other cases are more mysterious. It is interesting to further study 
those theories.  To give an example, let's consider theories $(A_2;Y_n,\ldots, Y_1;Y_0)$ with $Y_n=\ldots=Y_{n_1+1}=[2,1]$, $Y_{n_1}=\ldots=Y_1=[1,1,1]$, here we take $n_1>1$.
The 3d mirror of this theory is shown in figure. \ref{nonconformal}.  

\newpage
\subsection{Duality}
Let's summarize the classification of AD matter and SCFT with exact marginal deformations found in previous section.  Let's start with 6d $A_{N-1}$ $(2,0)$ theory on a sphere with the following irregular singularity:
\begin{equation}
\Phi={T_n\over z^n}+\ldots+{T_1\over z},
\end{equation}
and the eigenvalue degeneracy of $T_i$ is encoded by a sequence of Young Tableaux
\begin{equation}
Y_{n}\subset Y_{n-1}\subset \ldots \subset Y_1.
\end{equation}
One can add another regular singularly labeled by  $Y_0$ so it is still possible to define a SCFT.  Our theory might be denoted as $(A_{N-1}; Y_{n},\ldots, Y_{1}; Y_0)$. 
We have following classification of SCFT:
\begin{itemize}
\item  For $n\geq 4$, the irregular singularity has the structure $Y_n=Y_{n-1}=\ldots=Y_2=[n_1, n_2,\ldots, n_a]$. $Y_1$ and $Y_0$ are arbitrary. We label this class of theory as $\textbf{(p,1)}$ with $p=n-2\geq 2$. One of the reason for the labeling
is that the Coulomb branch operators have the common denominator $p+1$.  
\item For $n\leq 3$,  and there are two possibilities:
\begin{enumerate}
\item The irregular singularity has the structure $Y_3=Y_2=[n_1, n_2, \ldots, n_a]$, $Y_1$ arbitrary.  (Theory defined with an irregular singularity and a regular singularity can be engineered using a single irregular singularity using a different 6d $(2,0)$ theory \cite{Xie:2012hs}.).
We call them theory of class $\textbf{(1,1)}$. 
\item The irregular singularity has the structure $Y_2=[n_1, n_2, \ldots, n_a]$, and $Y_1$ and $Y_0$ are both arbitrary. We call them theory of class $\textbf{(0,1)}$. 
\end{enumerate}
\end{itemize}
For $n\geq 4$, the number of exact marginal deformations are $a-2$, and for $n\geq 3$, the number of exact marginal deformations are $a-3$.

\subsubsection{Theory of class $(0,1)$}
Interestingly, class ${\cal S}$ theory  defined on sphere is identified as class $(0,1)$ theory, which is engineered by the following data $(A_{N-1}; Y_2, Y_1; Y_0)$, with $Y_2=[n_1,\ldots, n_a]$.
The irregular singularity has the following form
\begin{equation}
\Phi={T_2\over z^2}+{T_1\over z}.
\end{equation}
There is also an extra regular singularity labeled by $Y_0$. This theory has $a-3$ exact marginal deformations. The 3d mirror of this theory is a star shaped quiver with 
a central node of rank $n_c$ \cite{Benini:2010uu}, see also figure. \ref{scft}.

$S$ duality of class ${\cal S}$ theory is best understood by representing SCFT by a sphere with marked points \cite{Gaiotto:2009we}: We represent above SCFT by a sphere 
with $a$ marked points, and assign a Young Tableaux of size $n_c$. The number of complex structure deformation of the punctured sphere is $a-3$ which is equal to the number of exact marginal 
deformation of  SCFT.  Different weakly coupled gauge theory descriptions are then identified as different pants decomposition of the same punctured sphere.
Moreover the matter contents in the gauge theory description are identified as the theory represented by three punctured spheres. Many aspects of these dualities have been studied in \cite{Gaiotto:2009hg,Tachikawa:2009rb,Nanopoulos:2009xe, Nanopoulos:2010ga,Chacaltana:2010ks,Tachikawa:2010vg,
Chacaltana:2011ze, Chacaltana:2012ch,Chacaltana:2013oka,Chacaltana:2014jba,Chacaltana:2014nya,Chacaltana:2015bna}. These results generalize S duality results found by Argyres and Seiberg \cite{Argyres:2007cn,Argyres:2007tq,Argyres:2010py}.

\subsubsection{Theory of class $(1,1)$}
Let's consider theory  $(A_{N-1};Y_3, Y_2, Y_1)$ with $Y_3=Y_2=[n_1, n_2,\ldots, n_a]$ and $Y_1$ arbitrary.  
The irregular singularity has the following form
\begin{equation}
\Phi={T_3\over z^3}+{T_2\over z^2}+{T_1\over z^1}.
\end{equation}
This theory has $a-3$ exact marginal deformations.  

\begin{figure}[h]
\centering
  \includegraphics[width=4in]{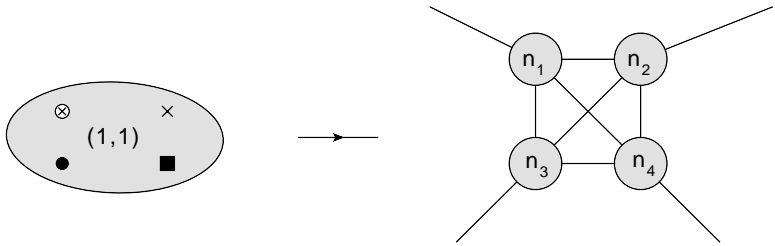}
  \caption{Left: A class (1,1) SCFT is represented by a sphere with marked points, and each marked point has a Young Tableaux with size $n_i$. Right: The 3d mirror of the theory shown on left, and 
  we omit the quiver tails which are determined by Young Tableaux.}
  \label{oneone}
\end{figure}

Motivated by  $S$ duality of class $(0,1)$ theory ( class ${\cal S}$ theory) and the result in \cite{Xie:2016uqq},
it might be also possible to represent our theory by a sphere with $a$ marked points: each marked point  has a rank $n_i$, and  a Young Tableaux $[m_1^{(i)}, \ldots, m_{r_i}^{(i)}]$ such 
that $\sum_{j=1}^{r_i} m_j^{(i)}=n_i$.  See figure. \ref{oneone} for illustration, notice that in this case the size of Young Tableaux is different for each marked point. 
 One nice consequence of this identification is that the number of exact marginal deformation matches with 
the complex structure moduli of punctured sphere, and  AD matter is represented by a three punctured sphere. 

With the above representation of our SCFT,  we would like to propose the following method of identifying the duality frames for class $(1,1)$ theories:

\begin{conjecture}
Different duality frames of class $(1,1)$ theory labeled by the data $(A_{N-1};Y_3, Y_2, Y_1)$ with $Y_3=Y_2=[n_1, n_2,\ldots, n_a]$  is identified with different degeneration limit of a sphere with $a$ marked points. 
\end{conjecture}

\begin{figure}[h]
\centering
  \includegraphics[width=5in]{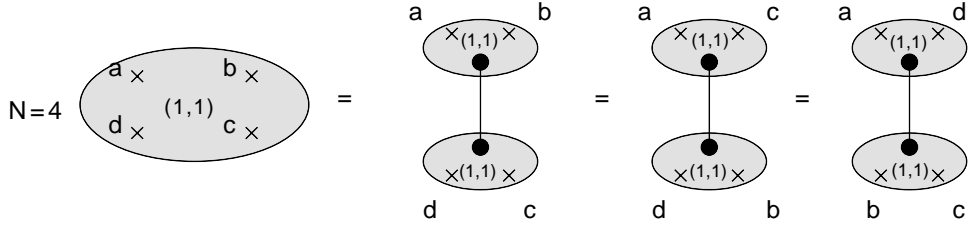}
  \caption{The three duality frames of $(A_3, A_3)$ theory. The cross marked point has Young Tableaux $[1]$. }
  \label{N1}
\end{figure}

\begin{figure}[h]
\centering
  \includegraphics[width=5in]{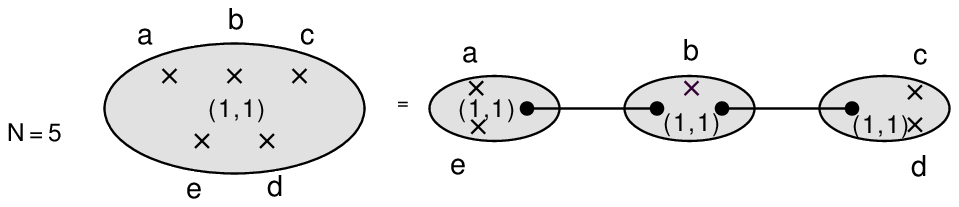}
  \caption{The  duality frames of $(A_4, A_4)$ theory. Different labelings for the marked points in the degeneration limit of the punctured sphere give different duality frames. The cross marked point has Young Tableaux $[1]$.}
  \label{N2}
\end{figure}

\begin{figure}[h]
\centering
  \includegraphics[width=5in]{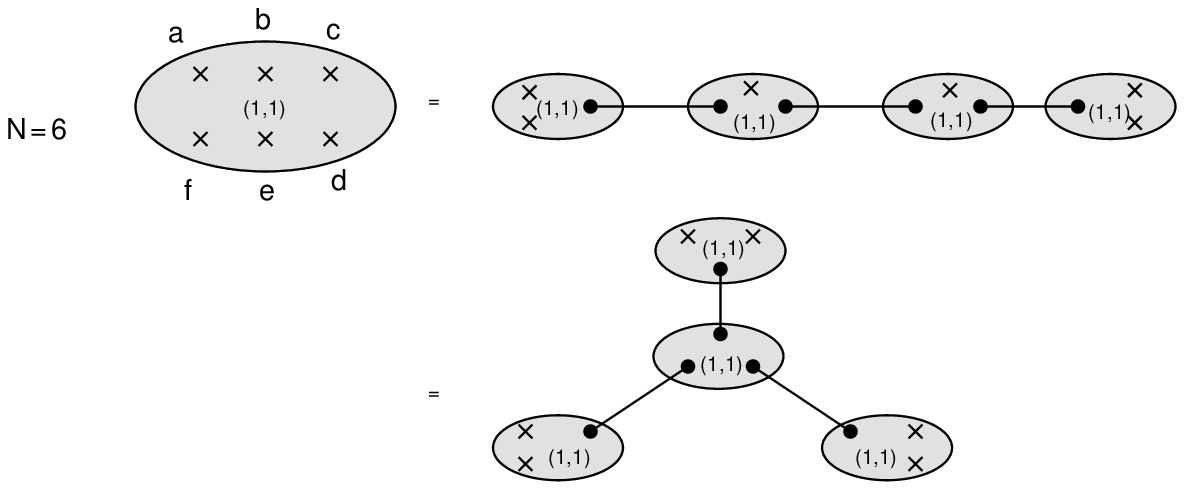}
  \caption{The  duality frames of $(A_5, A_5)$ theory. Different labelings for the marked points in the degeneration limit of the punctured sphere give different duality frames. The cross marked point has Young Tableaux $[1]$.}
  \label{N3}
\end{figure}

\textbf{Example}:  Let's consider the theory labeled by the data  $(A_{N-1};Y_3, Y_2, Y_1)$ with $Y_3=Y_2=Y_1=[1, \ldots, 1]$. This class of theory can also be engineered by Type IIB string theory on 
a 3-fold singularity $x_1^2+x_2^2+x_3^N+x_4^N=0$ (They are also called $(A_{N-1}, A_{N-1})$ theory \cite{Cecotti:2010fi}.). To study $S$ duality of this theory, we  represented it by a sphere with $N$ identical marked points, which is of the type $[1]$.  Different duality frames 
are identified as different degeneration limits of the same punctured sphere, see figure. \ref{N1}, \ref{N2}, \ref{N3}. 

\newpage

The field theory description for these theories can be read from the decompositions of the punctured sphere: 

N=4:~~
\begin{equation}
\xymatrix{
	&  &  1\ar@{-}[d]
	\\
	&  D_2SU(3) \ar@{-}[r]&  \textbf{SU(2)} \ar@{-}[r]&D_2(SU(3)) 
}
\end{equation}

N=5:~~
\begin{equation}
\xymatrix{
	&  D_2SU(3) \ar@{-}[r]&  \textbf{SU(2)} \ar@{-}[r]&D_2(SU(5))  \ar@{-}[r]&  \textbf{SU(2)} \ar@{-}[r]&D_2(SU(3)) 
}
\end{equation}

N=6:
\begin{equation}
\xymatrix{
	&  &  &&1\ar@{-}[d]
	\\
	&  D_2SU(3) \ar@{-}[r]&  \textbf{SU(2)} \ar@{-}[r]&D_2(SU(5)) \ar@{-}[r] & \textbf{SU(3)} \ar@{-}[r] &D_2(SU(5))   \ar@{-}[r]& \textbf{ SU(2)} \ar@{-}[r]&D_2(SU(3)) 
}
\end{equation}
\begin{equation}
\xymatrix{
	&  &  &D_2SU(3)\ar@{-}[d]
\\
	&  &  &\textbf{SU(2)}\ar@{-}[d]
	\\
	&  D_2SU(3) \ar@{-}[r]&  \textbf{SU(2)} \ar@{-}[r]&T \ar@{-}[r] &   \textbf{SU(2)} \ar@{-}[r]&D_2(SU(3)) 
}
\end{equation}

Let's explain how to find the duality frames for $N=4$.  In the degeneration limit, we get two three punctured spheres: two marked points are of the type $[1]$, 
while the third one coming from degeneration limit has the type $[1,1]$. This theory can be engineered using the irregular singularity $(A_3; [1,1,2],[1,1,2],[1,1,1,1])$. The 3d mirror of this 
theory is a bad quiver, and use the reduction procedure, we find that the interacting part is engineered by the irregular singularity $(A_2; [1,1,1],[1,1,1],[1,1,1])$.
This theory has a $SU(3)$ flavor symmetry, and we gauge a $SU(2)$ subgroup.
The extra free hyper on $SU(2)$ gauge group  is needed for the conformal invariance. More checks will be given later. 

\begin{figure}[h]
\centering
  \includegraphics[width=2.5in]{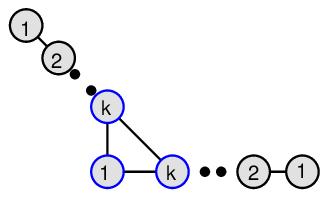}
  \caption{The 3d mirror for $D_2(SU(2k+1))$ theory which has $SU(2k+1)$ flavor symmetry.}
  \label{AD2}
\end{figure}

\begin{figure}[h]
\centering
  \includegraphics[width=1.5in]{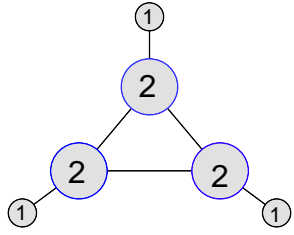}
  \caption{3d mirror for theory $T=(A_5;[2,2,2],[2,2,2],[1,1,1,1,1,1])$. }
  \label{TT}
\end{figure}

Let's explain more the AD matter used in the field theory description.
AD matter $D_2SU(2k+1)$ is engineered using the following irregular singularity 
\begin{equation}
Y_3=[k,k,1],~Y_2=[k,k,1],~Y_1=[1,\ldots,1].
\end{equation}
and the 3d mirror is shown in figure. \ref{AD2}.
This theory has global symmetry $SU(2k+1)$, and various data is listed in table. \ref{ADdata}.
\begin{table}[!htb]
\begin{center}
  \begin{tabular}{ |c|c|c|c|c|}
    \hline
     AD matter & Coulomb branch spectrum & a&c &$k_F$\\ \hline
     $D_2SU(2k+1)$ & $\{ {2k+1\over 2},~{2k-1\over 2},~\ldots,~{3\over2},~\underbrace{1,\ldots,1}_{2k} \}$& $\frac{7}{24} k (k+1)$ & $\frac{1}{3} k (k+1)$&${2k+1\over 2}$ \\ \hline
     $T$&$\{{3\over2},{5\over2},{5\over2},3\}$&${37\over12}$&${41\over12}$&$k_{SU(2)}={5\over2}$\\ \hline
  \end{tabular}
  \end{center}
  \caption{Various quantities for the AD matter $D_2SU(2k+1)$ and $T$ which appear in $S$ duality of $(A_{N-1}, A_{N-1})$ theory with $N\leq 6$.}
  \label{ADdata}
\end{table}
The theory $T$ can be engineered using six dimensional $A_5$ theory on the following irregular singularity 
\begin{equation}
Y_3=[2,2,2],~Y_2=[2,2,2],~Y_1=[1,1,1,1,1,1].
\end{equation}
This theory has flavor symmetry $SU(2)^3\times U(1)^2$. The 3d mirror is shown in figure. \ref{TT}, and various other data for this SCFT is listed in table. \ref{ADdata}. 

 Let's explain some checks for $S$ dualities we proposed for $(A_{N-1}, A_{N-1})$ theory:
\begin{itemize}
\item First of all, the sum of $\beta$ function is zero for all the gauge groups. For $N=4$, we have $\beta_{D_2(SU(3))}={3\over 2}$ and $\beta_{hyper}=1$ to the $SU(2)$ gauge group, so the sum of 
the contribution from various matter contents is 4, which cancels the contribution from the vector multiplet of $SU(2)$ gauge group.
\item The central charge $a$ and $c$ of the original theory is computed and listed in table. \ref{ADdata}, which can be derived from the formula in (\ref{central}). 
Let's do the computation for $N=4$ explicitly. Using gauge theory description, we find 
\begin{align}
&a_{total}=a_{vec}+a_{hyper}+2a_{D_2(SU(3))}={5\over 8}+{1\over 12}+2*{7\over 12}={15\over8}, \nonumber\\
&c_{total}=c_{vec}+c_{hyper}+2c_{D_2(SU(3))}=  {1\over 2}   +{1\over 6}+2*{2\over3}=2.
\end{align}
The answer is the same as the central charge of the original theory listed in table. \ref{aNaN}.
\item The Coulomb branch spectrum of the original theory is listed in table. \ref{aNaN}, which can be either computed from the method reviewed in previous section or the method 
using singularity theory \cite{Xie:2015rpa}. We can compute the coulomb branch spectrum using the gauge theory description, 
and explicitly verify that it agrees with the result in table. \ref{aNaN}. For $N=4$, each $D_2 SU(3)$ matter sector contributes a coulomb branch operator with scaling dimension ${3\over 2}$ and 
a mass parameter, and $SU(2)$ gauge group contributes a dimension two operator, finally the free hyper contributes a mass parameter, so the Coulomb branch spectrum 
is $(1,1,1,{3\over 2}, {3\over 2}, 2)$, which matches with the Coulomb branch spectrum of original theory. 
\end{itemize}

\begin{table}[!htb]
\begin{center}
  \begin{tabular}{ |c|c|c|c|}
    \hline
     $N$& Coulomb branch spectrum & a&c \\ \hline
     4 & $\left\{1,1,1,\frac{3}{2},\frac{3}{2},2\right\}$& $\frac{15}{8}$ & $2$ \\ \hline
     5 & $\left\{1,1,1,1,\frac{3}{2},\frac{3}{2},\frac{3}{2},2,2,\frac{5}{2}\right\}$ & $\frac{25}{6}$ & $\frac{13}{3}$ \\ \hline
     6&$\left\{1,1,1,1,1,\frac{3}{2},\frac{3}{2},\frac{3}{2},\frac{3}{2},2,2,2,\frac{5}{2},\frac{5}{2},3\right\}$ & $\frac{185}{24}$ &$\frac{95}{12}$ \\ \hline
  \end{tabular}
  \end{center}
  \caption{ Various quantities for $(A_{N-1}, A_{N-1})$ theories.}
  \label{aNaN}
  \label{d}
\end{table}

\begin{figure}[h]
\centering
  \includegraphics[width=6in]{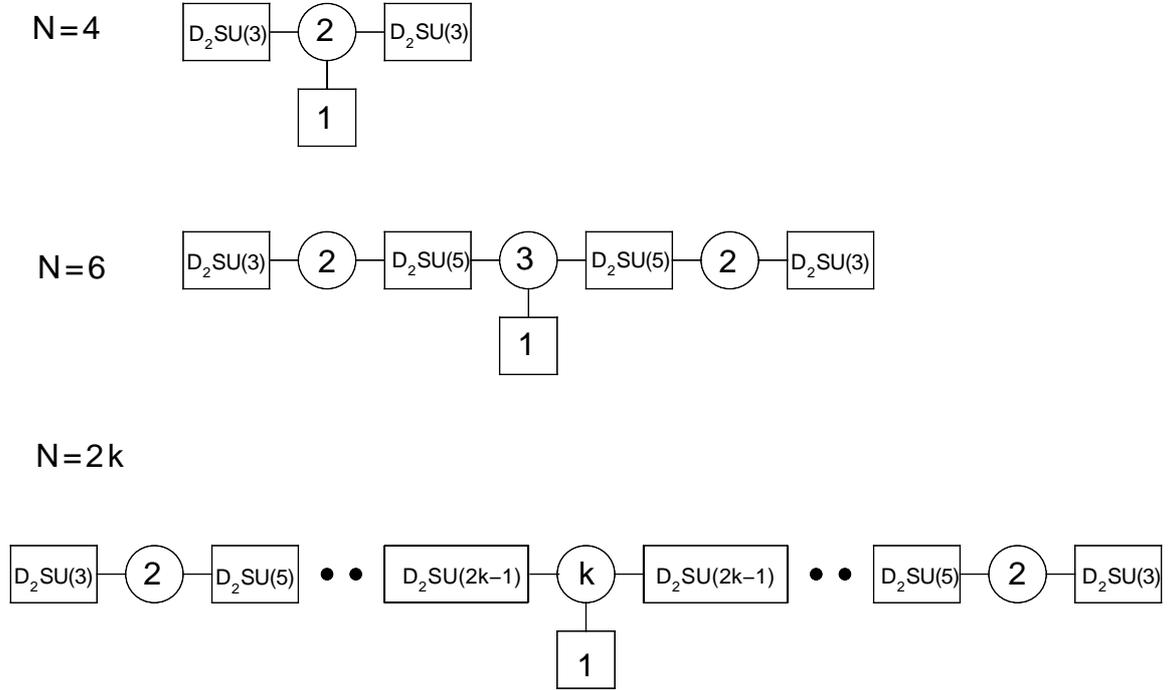}
  \caption{One duality frame for $(A_{N-1}, A_{N-1})$ theory with even $N$.}
  \label{aNaNeven}
\end{figure}

\begin{figure}[h]
\centering
  \includegraphics[width=6in]{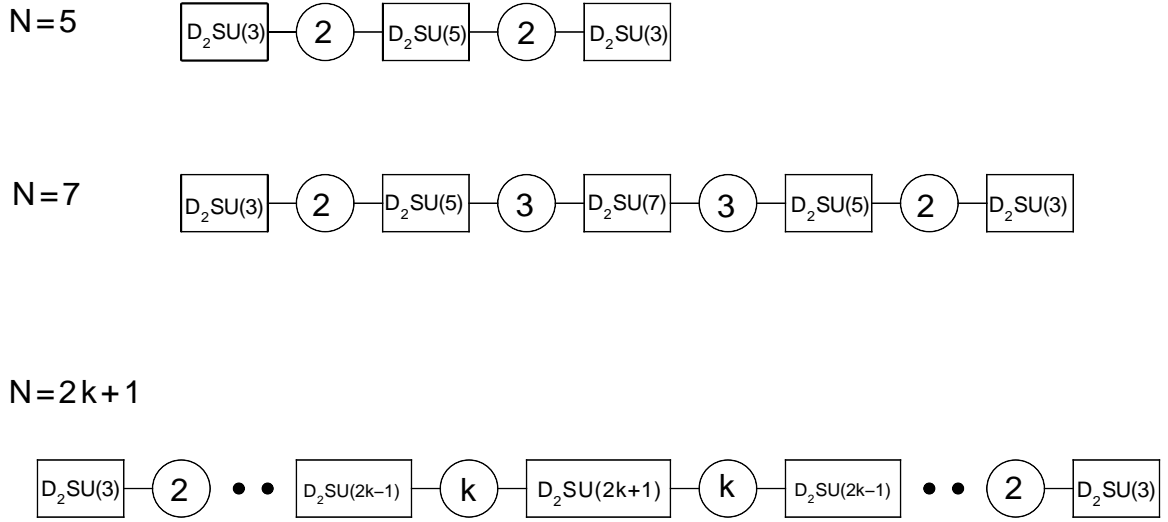}
  \caption{One duality frame for $(A_{N-1}, A_{N-1})$ theory with odd $N$.}
  \label{aNaNodd}
\end{figure}

For general $N$, the matter contents can be found as follows: a  
tube in the degeneration limit separates the original marked points into two sets with number $n_L$ and $n_R$ respectively;  the new marked point in the degeneration limit is of the type $\underbrace{[1,\ldots, 1]}_{min(n_L, n_R)}$.
 We also need to use the reduction procedure on 3d mirror to get the interacting part. Sometimes, free fundamental matter is also needed to ensure conformal invariance.
A duality frame where the interacting matter systems  are $D_2 SU(2k+1)$ theory is shown in figure. \ref{aNaNeven} and figure. \ref{aNaNodd}. 

\newpage
\subsubsection{Theory of class $(p,1)$}
Let's now consider the SCFT $(A_{N-1}; Y_{n}, \ldots, Y_1; Y_0)$ with $Y_n=\ldots=Y_2=[n_1,\ldots, n_a]$, $n\geq 4$. The irregular singularity has the following form:
\begin{equation}
\Phi={T_n\over z^n}+{T_{n-1}\over z^{n-1}}+\ldots+{T_1\over z^1}.
\end{equation}
The number of exact marginal 
deformations is $a-2$. This is the theory of class $(p,1)$ with $p=n-2$.

\begin{figure}[h]
\centering
  \includegraphics[width=4in]{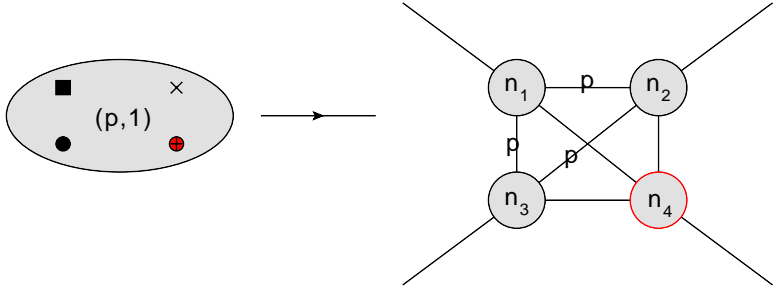}
  \caption{The punctured sphere for class $(p,1)$ theory. The red marked point representing the regular puncture. The 3d mirror for this theory is shown on  the right hand side, and 
  the quiver tail is represented by a straight line whose detailed form is determined by the Young Tableaux of the marked point.}
  \label{ponemarked}
\end{figure}

We now represent our theory by a sphere with $a+1$ marked points, and we have two kinds of marked points: one  represents the data of $Y_i,~i\geq2$ and $Y_1$: there 
is a Young Tableaux attached to each marked point and its total size is $n_i$ (one of the column data of the Young Tableaux $Y_2$), and we label it as black puncture;
We also need an extra Young Tableaux $Y_0$ with size $N$, and we need this marked point even if $Y_0$ is trivial (the Young Tableaux is the type $[N]$). See figure. \ref{ponemarked}. 
We label it as a red puncture, and the size of its Young Tableaux is the sum of size of 
all the black punctures. The AD matter is represented by a three punctured sphere with two black puncture and one red puncture.

Different duality frames are represented by different degeneration limits of the same punctured sphere. In the degeneration limit, one get opposite type of marked points on two degenerated pieces, 
so that each punctured three sphere in the degeneration limit  has only one red  puncture and two black punctures. 

\begin{figure}[h]
\centering
  \includegraphics[width=4.5in]{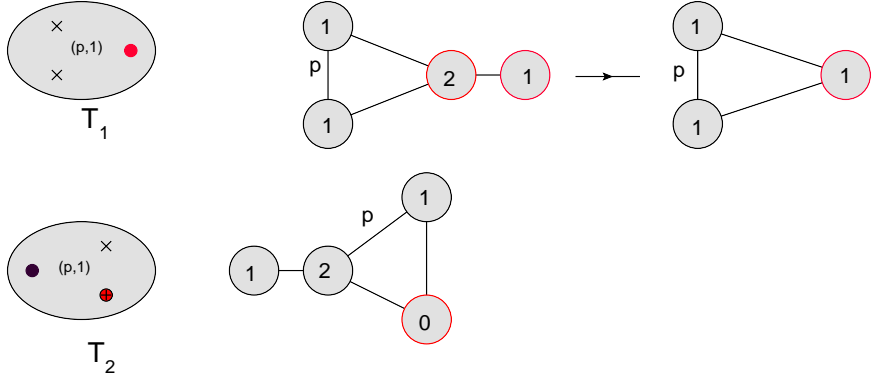}
  \caption{The 3d mirror for two matter systems appearing in the duality frame of $(A_2, A_{2p-1})$ theory.}
  \label{poneduality2}
\end{figure}

\textbf{Example}
Let's  consider SCFT $(A_{N-1}; Y_{n}, \ldots, Y_1; Y_0)$ with $Y_n=\ldots=Y_1=[1,\ldots, 1]$ and $Y_0=[N]$. 
The same theory can be defined by the singularity $f=z_0^2+z_1^2+z_2^N+z_3^{(n-2)N}$. Now we represent the theory by a sphere with 
$N+1$ marked points , and the first $N$ marked points are the same and of type $[1]$, while the extra marked point represents the trivial regular puncture with Young Tableaux $[N]$. 
$S$ duality is interpreted as different degeneration limits of this punctured sphere.  To find out the weakly coupled gauge theory description, we 
need to figure out the new puncture in the degeneration limit.  Let's write down the field theory for $N=3$ case ( the general case 
could be worked out following the method used in \cite{Nanopoulos:2010ga}). The new puncture in the degeneration limit is of the type $[1,1]$, so the 3d mirror of two matter systems 
are shown in figure. \ref{poneduality2}, and the reduced interacting matter system is also shown there from which we can read off the M5 brane configuration.  The filed theory description is then
\begin{equation}
T_1-SU(2)-T_2.
\end{equation}
The two matter systems are engineered as follows
\begin{align}
&T_1:~~(A_1; Y_n,\ldots, Y_1; Y_0),~~Y_n=\ldots=Y_1=Y_0=[1,1]  \nonumber\\
&T_2:~~(A_2; Y_n,\ldots, Y_1; Y_0),~~Y_n=\ldots=Y_2=[2,1],~Y_1=[1,1,1],~Y_0=[3].
\end{align}
$T_1$ has flavor symmetry $SU(2)_1\times U(1)$ with $k_{SU(2)_1}=2+{1\over p+1}$;  $T_2$ also has flavor symmetry $SU(2)_2\times U(1)$ with $k_{SU(2)_2}=1+{p\over p+1}$.
So the contribution of matter system $T_1$ and $T_2$ to $SU(2)$ gauge group is four, which is needed for the conformal invariance. The interested reader 
can check that the central charges and Coulomb branch spectrum are the same for the full theory and the field theory description.

\begin{figure}[h]
\centering
  \includegraphics[width=4in]{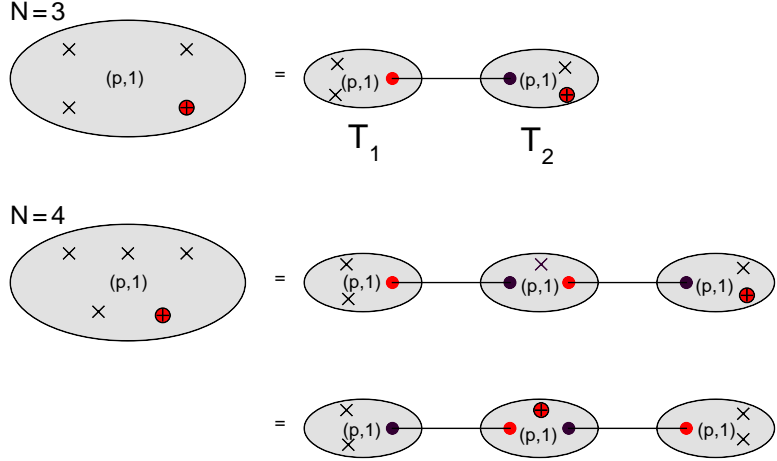}
  \caption{Different duality frames for $(A_{N-1}, A_{pN-1})$ theory. The black marked point has Young Tableaux $[1]$, the red marked point has Young Tableaux $[N]$. We do not write down the label for each black marked point, and the permutation of these labels will give different duality frames. }
\end{figure}

\newpage
\section{Theory of class $(p,q)$}
Let's now consider the following class of SCFTs: the Coulomb branch spectrum takes the form $\Delta(u)= {p i\over p+q}$ or $\Delta (u)= {q i\over p+q}$, with $(p,q)=1$ and $i$ integers.
We call them theories of class $(p,q)$ (we also impose the condition $q> 0$).  Using irregular and regular punctures reviewed in section II, we have two interesting family of theories in class $(p,q)$, and 
we call them class A and class B theory. The corresponding Newton diagram is listed in figure. \ref{pqnewton}, and the number of exact marginal deformations is $r-1$.
\begin{figure}[h]
\centering
  \includegraphics[width=3in]{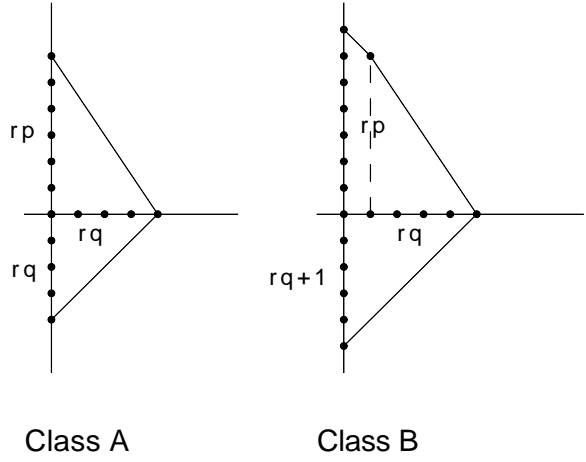}
  \caption{Newton polygon for two family of class $(p,q)$ theories.}
  \label{pqnewton}
\end{figure}

Let's consider class A theory: start with 6d  $A_{rq -1}$ $(2,0)$ theory and compactify it on a sphere with following irregular singularity:
\begin{equation}
\Phi={T_{p+q+1}\over z^{2+{p\over q}}}+{T_{p+q}\over z^{2+{p-1\over q}}}+\ldots+{T_{1}\over z}.
\end{equation}
The diagonal matrix $T_i$ takes the following form
\begin{equation}
T_i=\text{diag}(a_1 A_i, a_2 A_i,\ldots, a_r A_i),~~a_i\neq a_j.
\label{pqirre}
\end{equation}
Here $A_i$ is a fixed q dimensional diagonal matrix. We can add another regular singularity,  sand the theory is 
still in class $(p,q)$. In this case $p$ could take negative value with the range $ p> -q$.  
The SW curve at the SCFT point is 
\begin{equation}
x^{rq}+z^{rp}=0 \rightarrow [x]={p\over p+q},~~[z]={q\over p+q}.
\end{equation}
One can compute the full Coulomb branch spectrum by considering the deformation of above curve, and they can also be read from 
the under-diagram monomials of the Newton polygon.  Similarly, the description of irregular singularity and the Coulomb branch spectrum 
can be found for type B theory.

We now consider the degeneration of irregular singularity (\ref{pqirre}), and  they are classified by a sequence of Young Tableaux 
\begin{equation}
Y_{p+q+1}\subset Y_{p+q} \subset \ldots \subset Y_1.
\end{equation}
Here $Y_i$ is a  Young Tableaux $[n_1,\ldots, n_{r_i}]$ with $\sum n_j= r$, and $n_i$ denotes the degeneracy of $a_i$ in formula (\ref{pqirre}).  The Coulomb branch spectrum can be found as follows: 
Define the covering coordinate $w$ using formula $z=w^q$, and $dz=qw^{q-1} dw$, then the irregular singularity has the form (the Higgs field is a one form $\Phi(z) dz$):
\begin{align}
& \Phi(w) dw=({T_{p+q+1}\over w^{q+p+1}}+{T_{p+q}\over w^{q+p}}+\ldots+{T_{1}\over w})dw. \nonumber\\
& Y_{p+q+1}^{'}\subset Y_{p+q}^{'} \subset \ldots \subset Y_1^{'}.
\end{align}
and Young tableaux $Y_i^{'}=q Y_i= [\underbrace{n_1,\ldots, n_1}_{q}, \ldots, \underbrace{ n_{r_i},\ldots, n_{r_i}}_{q}]$. 
We can now use the formula (\ref{spec}) to find the maximal power $d_i$ of  the form $\omega^{d_i} x^{ rq -i}$ which will give 
Coulomb branch operators. We can change back to $z$ coordinate by noticing that $\omega^{d_i}$ transforms as a degree $i$ differential on Riemann surface, i.e.
\begin{equation}
\omega^{d_i} (dw)^i=\omega^{d_i} ({dz\over w^{q-1}})^i=z^{d_i-(q-1)i\over q} (dz)^i.
\end{equation}
The maximal power of $z^{d_{i}^{'} }x^{rq -i}$  is then 
\begin{equation}
d_i^{'}=[{d_i-(q-1)i\over q}]. 
\label{origina}
\end{equation}
Here square bracket means taking the integral part of number inside. 

\textbf{Example}: Let's consider  $A_3$ $(2,0)$ theory on a sphere with following irregular singularity 
\begin{equation}
\Phi={T_6\over z^{2+{3\over 2}}}+{T_5\over z^{3}}+\ldots+{T_1\over z}.
\end{equation}
We have $(p,q)=(3,2)$, and the Young Tableaux is $Y_6=\ldots=Y_2=[2]$, and $Y_1=[1,1]$.  Let's choosing the covering coordinate $x=\omega^2$, and the Higgs field becomes
\begin{equation}
\Phi={T_6\over w^6}+\ldots+{T_1\over w}.
\end{equation}
The Young Tableaux is $Y_6=\ldots=Y_2=[2,2]$ and $Y_1=[1,1,1,1]$. We find 
\begin{equation}
d_i=(0, 2, 1, 5)
\end{equation}
and using formula (\ref{origina}):
\begin{equation}
d_i^{'}=(-1,0,-1,0)
\end{equation}
and the SW curve with independent Coulomb branch operators looks like 
\begin{equation}
x^4+ [u_1]x+z^6+[u_2]=0,
\end{equation}
and the scaling dimensions are $u_2={12\over 5},~ u_1={6\over 5}$. This theory has flavor symmetry $SU(2)$, and
we identify it as the rank two $H_0$ theory \cite{Aharony:2007dj}.

We now classify the configuration which defines a SCFT. Let's consider SW curve of our theory
\begin{equation}
x^{rq}+\sum_{i=2}^{rq} \phi_i x^{rq -i}=0.
\end{equation}
Let's first consider the case $p>q\geq 2$, then
the dimension two operators appear as coefficient of following monomial
\begin{equation}
z^{b(j)}x^{rq -j},~~b(j)=-2+{(j-2)p\over q},~~j=2+k q,~~k=1,\ldots, r-1.
\label{dtwo}
\end{equation}
Now let's assume $Y_{p+q+1}=[n_1, \ldots, n_a]$ with $\sum_{j=1}^a a_j=r$, then there are at most $a-1$ exact marginal deformations. 
Let's first analyze the differential $\phi_j,~j=2+kq,~~k=1,\ldots, a-1$, and the dimension two operators in those 
differential are preserved as there is no reduction for these differentials, so there is at least $a-1$ dimension two operators which 
match with the number of exact marginal deformations. 
Let's next consider differential $\phi_{2+aq}$, and using formula \ref{origina}, we have 
\begin{equation}
d_{2+a q}=[{ n_1(2+aq-2)+n_2(2+aq-1)-(q+1)(2+a q)\over q}]=[-2+ap+{n_2-2\over q}]
\end{equation}
here $n_1$ (resp. $n_2$) denote the number of Young Tableaux where $aq+2$ is in second row (resp. first row),  we used the fact that $n_1+n_2=p+q+1$. 
To have a SCFT, we need to have $d_{2+aq}<b(2+aq)$ (see formula (\ref{dtwo})), so $n_2=0$ or $n_2=1$.   
The conclusion is that the constraint on the Young Tableaux is again
\begin{equation}
Y_{p+q+1}=\ldots=Y_2,~~Y_1~\text{arbitrary}
\label{generalre}
\end{equation} 
Similar analysis  can be done for the case $-q+1\leq p<q$, and the result is the same as (\ref{generalre}). 

 The AD matter (for $q\geq 2$) (SCFT without exact marginal deformations) is now classified as follows:
 \begin{itemize}
 \item If $p\neq 1$,  AD matter is classified by the Young Tableaux $Y_{p+q+1}=\ldots=Y_2=[r]$, and $Y_1$, $Y_0$ are arbitrary.
\item If $p=1$ and $q$ arbitrary, we have 
\begin{equation}
b(2+k q)=-2+k,~~k=1,\ldots, r-1. 
\end{equation}
If $r=1$, one get AD matter. If $r=2$, and there is no regular singularity, then $b(2+q)=-1$, so there is no dimension two operator in Coulomb branch spectrum as 
the exponent of $z$ variable is non-negative due to the absence of regular singularity; and in this case, the configuration with $Y_{p+q+1}=[1,1]$ describes AD matter, which 
is nothing but $(A_{q-1}, A_1)$ theory.

\end{itemize}

\subsection{Duality}
To study duality, we would like to represent our theory by a sphere with marked points and identify the exact marginal deformations with 
the complex structure moduli of the sphere.  Let's start with class $A$ theory, and there are a total of $r-1$ exact marginal deformations. Comparing 
with the class $(p,1)$ theory, it is tempting to represent the irregular singularity by $r$ marked points which reflect $r$ block of the form of irregular singularity.  
We can add a further marked point representing the regular singularity (the label could be zero to represent the case where the regular
singularity is trivial: $Y_0=[N]$.). Now there are a total of $r+1$ marked points on sphere, and the number of complex structure moduli is $r-2$, which is 
one less than the number of exact marginal deformation. 
It appears that we still miss a marked point! 

The solution is to look at class $B$ theory, for this class of theory,  we have $r$ marked point (we call them black type) from the one compact segment of boundary of 
Newton polygon, and one marked point (we call them red type) from regular singularity.  We need to add another marked point (blue type) with rank one representing the second compact segment of 
Newton polygon, see figure. \ref{pqnewton}. So there is a total of $r+2$ marked points on sphere, and the number of complex structure moduli matches with 
the number of exact marginal deformations. 

Here comes the representation for class $A$ theory:  there is an extra blue type marked point whose rank is zero.  Generally, we have 
the following representation for our SCFT: there are $r$ black marked point, one red marked point, and one blue marked point. We also 
attach a quiver tail to each black and red marked point, and for the blue marked point, only a $U(1)$ flavor symmetry is found.

\begin{figure}[h]
\centering
  \includegraphics[width=5in]{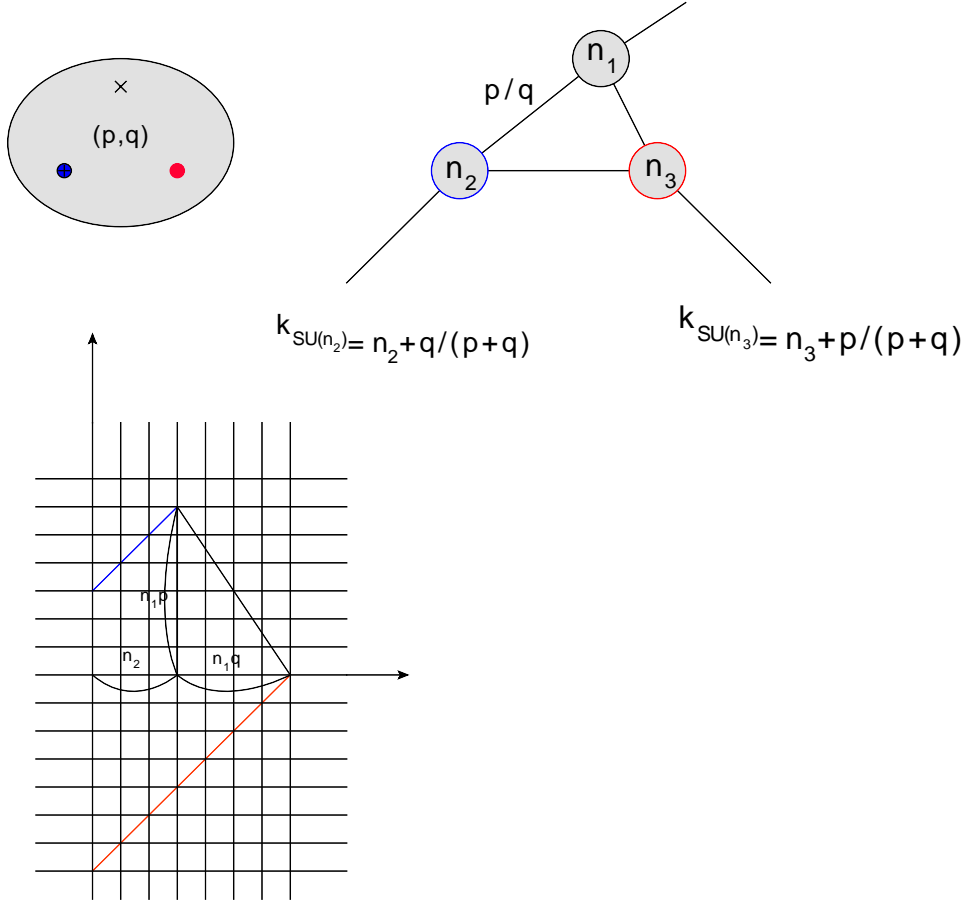}
  \caption{Top Left: The AD matter in class $(p,q)$ theories is represented by a sphere with one black marked point, one red and one blue marked point. Top right: A psedoquiver representation for 
  the AD matter, and this is not a real quiver since the quiver arrow is fractional. Bottom: Newton polygon for the AD matter and one can read off the Coulomb branch spectrum from the lattice points 
  under the boundary of Newton polygon. }
  \label{pqnewton}
\end{figure}

Using the classification of AD matter in last subsection, we see that AD matter is represented by a three sphere with three marked points.
\begin{itemize}
\item if $p\neq 1$, AD matter is represented by a sphere with one black, one blue and one blue marked point. 
Notice that we only have $U(1)$ flavor symmetry for blue marked point. 
\item if $p=1$, there are two kinds of AD matter: a): sphere with a black, red, and blue marked point; b): sphere
with two black, and one blue marked point. 
\end{itemize}

Our later study suggest that in general we need a theory whose blue 
marked point has general flavor symmetry, and the Coulomb branch spectrum might be found from a Newton polygon shown in figure. \ref{pqnewton}.
It is not obvious where this theory sits in our M5 brane configurations, and it would be interesting to find some stringy constructions. 

The duality is again interpreted as the degeneration limit of the punctured Riemann surface. However,  not all the degeneration limit seems possible as there is 
only one type of AD matter which  requires three different marked points (for $p\neq 1$ and $q\neq 1$.).

\begin{figure}[h]
\centering
  \includegraphics[width=5in]{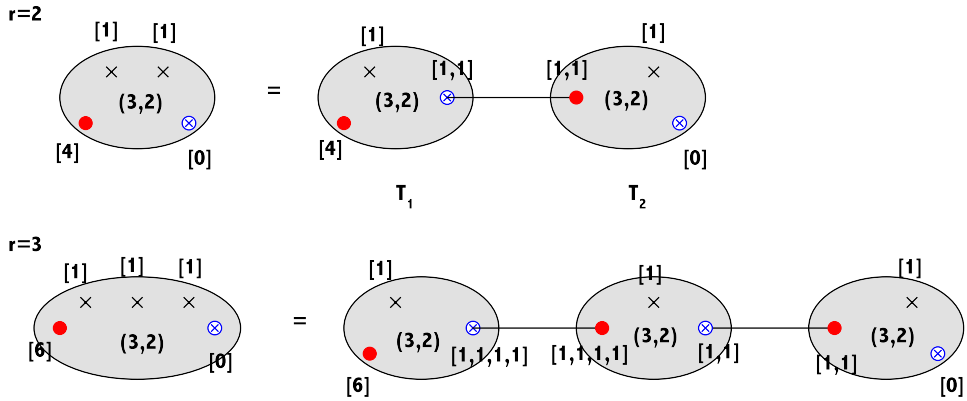}
  \caption{Duality frames for $(A_{2r-1}, A_{3r-1})$ theory}
  \label{pqduality}
\end{figure}

\begin{figure}[h]
\centering
  \includegraphics[width=3in]{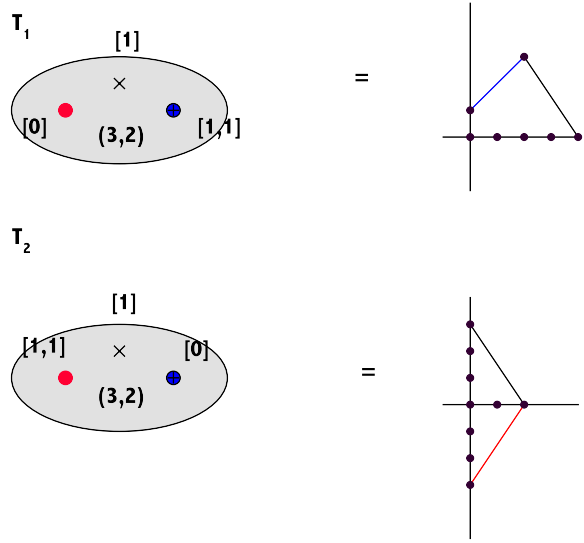}
  \caption{Two AD matter in the field theory limit of the theory engineered using 3-fold singularity $x_1^2+x_2^2+x_3^4+x_4^6=0$. The Coulomb branch 
  operators can be found from the monomials under the corresponding Newton polygon.}
  \label{pqduality1}
\end{figure}

\textbf{Example}
Let's take $p=3$ and $q=2$, consider the SCFT which is defined by the following irregular singularity 
\begin{equation}
\Phi={T_6\over z^{2+{3\over2}}}+\ldots+{T_1\over z}.
\end{equation}
Here $T_6$ is a $2r\times 2r$ diagonal matrix.  
This theory can also be engineered by the 3-fold singularity  $x_1^2+x_2^2+x_3^{2r}+x_4^{3r}=0$ (they are called $(A_{2r-1}, A_{3r-1})$ theory.).  
According to our general rule, it can be represented by a sphere with $r$ black marked point of type $[1]$, and 
one red marked point of type $[0]$, and one red marked point of type $[2r]$. 
S duality is interpreted as different degeneration limit of this punctured sphere. 
Let's take $r=2$ and $r=3$ for illustration, see figure. \ref{pqduality}.  For example, if $r=2$, the field theory looks like 
\begin{equation}
T_1-SU(2)-T_2.
\end{equation}
Here $T_1$ and $T_2$ are shown in figure. \ref{pqduality1}. The interested reader can check that the Coulomb branch spectrum, central charges of the gauge theory
description match with original theory. 

\newpage
\section{SCFT built from gauging AD matter}
We have studied $S$ duality of those SCFTs  which admit constructions using 6d $A_{N-1}$ $(2,0)$ theory. 
More generally, one could just gauge various AD matters to form new SCFTs, and some of them 
might not have string theory or M5 brane description. The theory space seems pretty large. 

In previous sections, all the duality frames of our SCFT is a linear quiver in the sense that a gauge group is coupled to only two AD matters. In general, 
We could also build D type and E type quiver, namely there would be gauge group coupled with 
more than two AD matter systems.  Some examples are studied in \cite{DelZotto:2015rca,Wang:2016yha}. It would be interesting 
to perform a full classification of such new theories. 

For class ${\cal S}$ theory (class $(0,1)$ theory), we 
can build theories represented by higher genus Riemann surface \cite{Gaiotto:2009we}. The basic field theory interpretation is following:
one can gauge two $SU(N)$ flavor symmetries of a single $T_N$ theory to form a SCFT with a $SU(N)$ gauge group which is conformal.  One might wonder whether this higher genus version is possible for our general class $(p,q)$ theories. 
 However, it turns out that such higher genus generalization for general class $(p,q)$ theories is not possible. 
The reason is following: our $(p,q)$ AD matter has flavor symmetry $SU(n_1)\times SU(n_2)\times SU(n_3)$, here $SU(n_3)$ comes from regular singularity, and the flavor 
central charge is 
\begin{equation}
k_{SU(n_2)}=n_2+{q\over p+q},~~k_{SU(n_3)} =n_3+{p\over p+q};~~q> 0
\end{equation}
Here $SU(n_3)$ flavor symmetry can be enhanced to $SU(n_3+1)$ if $n_1q+n_2 =n_3+1$, and  $SU(n_2)$ flavor 
symmetry can be enhanced to $SU(n_2+1)$ if $n_3+n_1{p\over q}=n_2+1$.
To get a conformal gauging, we can try following:
\begin{enumerate}
\item We take $n_2=n_3+1$ and require the enhancement of $SU(n_3)$ to $SU(n_3+1)$, so the flavor symmetry of 
AD matter has a subgroup $SU(n_3+1)\times SU(n_3+1)$. We can gauge two $SU(n_3+1)$ flavor groups to 
get a conformal gauging, and we have  following two equations:
\begin{equation}
n_1q+n_2  =n_3+1,~~n_2=n_3+1,~~q> 0,
\end{equation}
but there is no solution, so we could not form a SCFT through gauging a single AD matter using this method.

\item Instead, we take $n_2=n_3-1$ and require the enhancement of $SU(n_2)$ to $SU(n_2+1)$, so we have following two equations:
\begin{equation}
n_3+n_1{p\over q}  =n_2+1,~~n_2=n_3-1,~~q\neq 0,
\end{equation}
This is only possible if $p=0$, and then $q=1$ by our convention. We get class $(0,1)$ theory which is just class ${\cal S}$ theory. 
\end{enumerate}

\section{Conclusion}
We have found a  $(p,q)$ generalization of $\mathcal{N}=2$ class ${\cal S}$ theory, and its $S$ duality behavior is found as follows: these
theories can be represented by a sphere with marked points, and $S$ duality is interpreted as different 
degeneration limits of the same punctured sphere.  Unlike class ${\cal S}$ theory, the punctured sphere in our 
case does not arise geometrically. It would be very interesting to find a stringy explanation of our result. 

It is surprising that $S$ duality of those more general AD theories can be interpreted in a similar way as 
class ${\cal S}$ theory. We hope that our findings  could help us understand better the still mysterious 
$S$ duality of supersymmetric quantum field theory.  More details such as the duality group deserve further study, 
see also \cite{Cecotti:2015hca,Caorsi:2016ebt}.

A natural generalization of our result would be to consider AD theories engineered from D and E type 6d $(2,0)$ theories \cite{Wang:2015mra}. 
The basic ides such as representing a theory by a punctured sphere should be similar, and it is interesting 
to work out the details. 

Our initial goal is to understand S duality of theories engineered using Type IIB string theory on 
a 3-fold singularity \cite{Shapere:1999xr,Xie:2015rpa,Chen:2016bzh,Wang:2016yha}. We only solved S duality for a very small subset of these theories:  
theories engineered using following singularities: $x_1^2+x_2^2+x_3^N+z^k=0$ and $x_1^2+x_2^2+x_3^N+x_3z^k=0$.
It seems that new methods are needed to understand $S$ duality of these general class of theories. 

It is interesting to explore the implication of our $S$ duality result on the computation of physical observables such 
as $S^4$ partition function, superconformal index, etc.

\section*{Acknowledgements}
The work of S.T Yau is supported by  NSF grant  DMS-1159412, NSF grant PHY-
0937443, and NSF grant DMS-0804454.  
The work of DX is supported by Center for Mathematical Sciences and Applications at Harvard University, and in part by the Fundamental Laws Initiative of
the Center for the Fundamental Laws of Nature, Harvard University.

\bibliographystyle{utphys}

\bibliography{ref}

\providecommand{\href}[2]{#2}\begingroup\raggedright\begin{thebibliography}{10}

\bibitem{Montonen:1977sn}
C.~Montonen and D.~I. Olive, ``{Magnetic Monopoles as Gauge Particles?},''
\href{http://dx.doi.org/10.1016/0370-2693(77)90076-4}{{\em Phys. Lett.} {\bf
  B72} (1977)  117}.

\bibitem{Seiberg:1994aj}
N.~Seiberg and E.~Witten, ``{Monopoles, duality and chiral symmetry breaking in
  N=2 supersymmetric QCD},''
  \href{http://dx.doi.org/10.1016/0550-3213(94)90214-3}{{\em Nucl. Phys.} {\bf
  B431} (1994)  484--550},
\href{http://arxiv.org/abs/hep-th/9408099}{{\tt arXiv:hep-th/9408099
  [hep-th]}}.

\bibitem{Argyres:2007cn}
P.~C. Argyres and N.~Seiberg, ``{S-duality in N=2 supersymmetric gauge
  theories},'' \href{http://dx.doi.org/10.1088/1126-6708/2007/12/088}{{\em
  JHEP} {\bf 12} (2007)  088},
\href{http://arxiv.org/abs/0711.0054}{{\tt arXiv:0711.0054 [hep-th]}}.

\bibitem{Gaiotto:2009we}
D.~Gaiotto, ``{N=2 dualities},''
  \href{http://dx.doi.org/10.1007/JHEP08(2012)034}{{\em JHEP} {\bf 08} (2012)
  034},
\href{http://arxiv.org/abs/0904.2715}{{\tt arXiv:0904.2715 [hep-th]}}.

\bibitem{Argyres:1995jj}
P.~C. Argyres and M.~R. Douglas, ``{New phenomena in SU(3) supersymmetric gauge
  theory},'' \href{http://dx.doi.org/10.1016/0550-3213(95)00281-V}{{\em Nucl.
  Phys.} {\bf B448} (1995)  93--126},
\href{http://arxiv.org/abs/hep-th/9505062}{{\tt arXiv:hep-th/9505062
  [hep-th]}}.

\bibitem{Argyres:1995xn}
P.~C. Argyres, M.~R. Plesser, N.~Seiberg, and E.~Witten, ``{New N=2
  superconformal field theories in four-dimensions},''
  \href{http://dx.doi.org/10.1016/0550-3213(95)00671-0}{{\em Nucl. Phys.} {\bf
  B461} (1996)  71--84},
\href{http://arxiv.org/abs/hep-th/9511154}{{\tt arXiv:hep-th/9511154
  [hep-th]}}.

\bibitem{Buican:2014hfa}
M.~Buican, S.~Giacomelli, T.~Nishinaka, and C.~Papageorgakis,
  ``{Argyres-Douglas Theories and S-Duality},''
  \href{http://dx.doi.org/10.1007/JHEP02(2015)185}{{\em JHEP} {\bf 02} (2015)
  185},
\href{http://arxiv.org/abs/1411.6026}{{\tt arXiv:1411.6026 [hep-th]}}.

\bibitem{DelZotto:2015rca}
M.~Del~Zotto, C.~Vafa, and D.~Xie, ``{Geometric engineering, mirror symmetry
  and $ 6{\mathrm{d}}_{\left(1,0\right)}\to
  4{\mathrm{d}}_{\left(\mathcal{N}=2\right)} $},''
  \href{http://dx.doi.org/10.1007/JHEP11(2015)123}{{\em JHEP} {\bf 11} (2015)
  123},
\href{http://arxiv.org/abs/1504.08348}{{\tt arXiv:1504.08348 [hep-th]}}.

\bibitem{Xie:2016uqq}
D.~Xie and S.-T. Yau, ``{New N = 2 dualities},''
\href{http://arxiv.org/abs/1602.03529}{{\tt arXiv:1602.03529 [hep-th]}}.

\bibitem{Dolan:2002zh}
F.~A. Dolan and H.~Osborn, ``{On short and semi-short representations for
  four-dimensional superconformal symmetry},''
  \href{http://dx.doi.org/10.1016/S0003-4916(03)00074-5}{{\em Annals Phys.}
  {\bf 307} (2003)  41--89},
\href{http://arxiv.org/abs/hep-th/0209056}{{\tt arXiv:hep-th/0209056
  [hep-th]}}.

\bibitem{Argyres:2015ffa}
P.~Argyres, M.~Lotito, Y.~LŸ, and M.~Martone, ``{Geometric constraints on the
  space of N=2 SCFTs I: physical constraints on relevant deformations},''
\href{http://arxiv.org/abs/1505.04814}{{\tt arXiv:1505.04814 [hep-th]}}.

\bibitem{Nanopoulos:2009uw}
D.~Nanopoulos and D.~Xie, ``{Hitchin Equation, Singularity, and N=2
  Superconformal Field Theories},''
  \href{http://dx.doi.org/10.1007/JHEP03(2010)043}{{\em JHEP} {\bf 03} (2010)
  043},
\href{http://arxiv.org/abs/0911.1990}{{\tt arXiv:0911.1990 [hep-th]}}.

\bibitem{Chacaltana:2012zy}
O.~Chacaltana, J.~Distler, and Y.~Tachikawa, ``{Nilpotent orbits and
  codimension-two defects of 6d N=(2,0) theories},''
  \href{http://dx.doi.org/10.1142/S0217751X1340006X}{{\em Int.J.Mod.Phys.} {\bf
  A28} (2013)  1340006},
\href{http://arxiv.org/abs/1203.2930}{{\tt arXiv:1203.2930 [hep-th]}}.

\bibitem{collingwood1993nilpotent}
D.~H. Collingwood and W.~M. McGovern, {\em Nilpotent orbits in semisimple Lie
  algebra: an introduction}.
\newblock CRC Press, 1993.

\bibitem{Xie:2012hs}
D.~Xie, ``{General Argyres-Douglas Theory},''
  \href{http://dx.doi.org/10.1007/JHEP01(2013)100}{{\em JHEP} {\bf 01} (2013)
  100},
\href{http://arxiv.org/abs/1204.2270}{{\tt arXiv:1204.2270 [hep-th]}}.

\bibitem{Wang:2015mra}
Y.~Wang and D.~Xie, ``{Classification of Argyres-Douglas theories from M5
  branes},''
\href{http://arxiv.org/abs/1509.00847}{{\tt arXiv:1509.00847 [hep-th]}}.

\bibitem{reeder2012gradings}
M.~Reeder, P.~Levy, J.-K. Yu, and B.~H. Gross, ``Gradings of positive rank on
  simple lie algebras,'' {\em Transformation Groups} {\bf 17} (2012) no.~4,
  1123--1190.

\bibitem{gaiotto2009wall}
D.~Gaiotto, G.~W. Moore, and A.~Neitzke, ``Wall-crossing, hitchin systems, and
  the wkb approximation,'' {\em arXiv preprint arXiv:0907.3987} (2009)  .

\bibitem{Cecotti:2011rv}
S.~Cecotti and C.~Vafa, ``{Classification of complete N=2 supersymmetric
  theories in 4 dimensions},'' {\em Surveys in differential geometry} {\bf 18}
  (2013)  ,
\href{http://arxiv.org/abs/1103.5832}{{\tt arXiv:1103.5832 [hep-th]}}.

\bibitem{Bonelli:2011aa}
G.~Bonelli, K.~Maruyoshi, and A.~Tanzini, ``{Wild Quiver Gauge Theories},''
  \href{http://dx.doi.org/10.1007/JHEP02(2012)031}{{\em JHEP} {\bf 02} (2012)
  031},
\href{http://arxiv.org/abs/1112.1691}{{\tt arXiv:1112.1691 [hep-th]}}.

\bibitem{elashvili2013cyclic}
A.~Elashvili, V.~Kac, and E.~Vinberg, ``Cyclic elements in semisimple lie
  algebras,'' {\em Transformation Groups} {\bf 18} (2013) no.~1, 97--130.

\bibitem{Cecotti:2010fi}
S.~Cecotti, A.~Neitzke, and C.~Vafa, ``{R-Twisting and 4d/2d
  Correspondences},''
\href{http://arxiv.org/abs/1006.3435}{{\tt arXiv:1006.3435 [hep-th]}}.

\bibitem{Witten:2007td}
E.~Witten, ``{Gauge theory and wild ramification},''
\href{http://arxiv.org/abs/0710.0631}{{\tt arXiv:0710.0631 [hep-th]}}.

\bibitem{Intriligator:1996ex}
K.~A. Intriligator and N.~Seiberg, ``{Mirror symmetry in three-dimensional
  gauge theories},'' \href{http://dx.doi.org/10.1016/0370-2693(96)01088-X}{{\em
  Phys. Lett.} {\bf B387} (1996)  513--519},
\href{http://arxiv.org/abs/hep-th/9607207}{{\tt arXiv:hep-th/9607207
  [hep-th]}}.

\bibitem{Gaiotto:2008ak}
D.~Gaiotto and E.~Witten, ``{S-Duality of Boundary Conditions In N=4 Super
  Yang-Mills Theory},''
  \href{http://dx.doi.org/10.4310/ATMP.2009.v13.n3.a5}{{\em Adv. Theor. Math.
  Phys.} {\bf 13} (2009) no.~3, 721--896},
\href{http://arxiv.org/abs/0807.3720}{{\tt arXiv:0807.3720 [hep-th]}}.

\bibitem{Nanopoulos:2010bv}
D.~Nanopoulos and D.~Xie, ``{More Three Dimensional Mirror Pairs},''
  \href{http://dx.doi.org/10.1007/JHEP05(2011)071}{{\em JHEP} {\bf 05} (2011)
  071},
\href{http://arxiv.org/abs/1011.1911}{{\tt arXiv:1011.1911 [hep-th]}}.

\bibitem{boalch2008irregular}
P.~Boalch, ``Irregular connections and kac-moody root systems,'' {\em arXiv
  preprint arXiv:0806.1050} (2008)  .

\bibitem{Benini:2010uu}
F.~Benini, Y.~Tachikawa, and D.~Xie, ``{Mirrors of 3d Sicilian theories},''
  \href{http://dx.doi.org/10.1007/JHEP09(2010)063}{{\em JHEP} {\bf 09} (2010)
  063},
\href{http://arxiv.org/abs/1007.0992}{{\tt arXiv:1007.0992 [hep-th]}}.

\bibitem{Gaiotto:2009hg}
D.~Gaiotto, G.~W. Moore, and A.~Neitzke, ``{Wall-crossing, Hitchin Systems, and
  the WKB Approximation},''
\href{http://arxiv.org/abs/0907.3987}{{\tt arXiv:0907.3987 [hep-th]}}.

\bibitem{Tachikawa:2009rb}
Y.~Tachikawa, ``{Six-dimensional D(N) theory and four-dimensional SO-USp
  quivers},'' \href{http://dx.doi.org/10.1088/1126-6708/2009/07/067}{{\em JHEP}
  {\bf 07} (2009)  067},
\href{http://arxiv.org/abs/0905.4074}{{\tt arXiv:0905.4074 [hep-th]}}.

\bibitem{Nanopoulos:2009xe}
D.~Nanopoulos and D.~Xie, ``{N=2 SU Quiver with USP Ends or SU Ends with
  Antisymmetric Matter},''
  \href{http://dx.doi.org/10.1088/1126-6708/2009/08/108}{{\em JHEP} {\bf 08}
  (2009)  108},
\href{http://arxiv.org/abs/0907.1651}{{\tt arXiv:0907.1651 [hep-th]}}.

\bibitem{Nanopoulos:2010ga}
D.~Nanopoulos and D.~Xie, ``{$N=2$ Generalized Superconformal Quiver Gauge
  Theory},'' \href{http://dx.doi.org/10.1007/JHEP09(2012)127}{{\em JHEP} {\bf
  09} (2012)  127},
\href{http://arxiv.org/abs/1006.3486}{{\tt arXiv:1006.3486 [hep-th]}}.

\bibitem{Chacaltana:2010ks}
O.~Chacaltana and J.~Distler, ``{Tinkertoys for Gaiotto Duality},''
  \href{http://dx.doi.org/10.1007/JHEP11(2010)099}{{\em JHEP} {\bf 11} (2010)
  099},
\href{http://arxiv.org/abs/1008.5203}{{\tt arXiv:1008.5203 [hep-th]}}.

\bibitem{Tachikawa:2010vg}
Y.~Tachikawa, ``{N=2 S-duality via Outer-automorphism Twists},''
  \href{http://dx.doi.org/10.1088/1751-8113/44/18/182001}{{\em J. Phys.} {\bf
  A44} (2011)  182001},
\href{http://arxiv.org/abs/1009.0339}{{\tt arXiv:1009.0339 [hep-th]}}.

\bibitem{Chacaltana:2011ze}
O.~Chacaltana and J.~Distler, ``{Tinkertoys for the $D_N$ series},''
  \href{http://dx.doi.org/10.1007/JHEP02(2013)110}{{\em JHEP} {\bf 02} (2013)
  110},
\href{http://arxiv.org/abs/1106.5410}{{\tt arXiv:1106.5410 [hep-th]}}.

\bibitem{Chacaltana:2012ch}
O.~Chacaltana, J.~Distler, and Y.~Tachikawa, ``{Gaiotto duality for the twisted
  A$_{2N-1}$ series},'' \href{http://dx.doi.org/10.1007/JHEP05(2015)075}{{\em
  JHEP} {\bf 05} (2015)  075},
\href{http://arxiv.org/abs/1212.3952}{{\tt arXiv:1212.3952 [hep-th]}}.

\bibitem{Chacaltana:2013oka}
O.~Chacaltana, J.~Distler, and A.~Trimm, ``{Tinkertoys for the Twisted
  D-Series},''
\href{http://arxiv.org/abs/1309.2299}{{\tt arXiv:1309.2299 [hep-th]}}.

\bibitem{Chacaltana:2014jba}
O.~Chacaltana, J.~Distler, and A.~Trimm, ``{Tinkertoys for the E$_{6}$
  theory},'' \href{http://dx.doi.org/10.1007/JHEP09(2015)007}{{\em JHEP} {\bf
  09} (2015)  007},
\href{http://arxiv.org/abs/1403.4604}{{\tt arXiv:1403.4604 [hep-th]}}.

\bibitem{Chacaltana:2014nya}
O.~Chacaltana, J.~Distler, and A.~Trimm, ``{A Family of $4D$ $\mathcal{N}=2$
  Interacting SCFTs from the Twisted $A_{2N}$ Series},''
\href{http://arxiv.org/abs/1412.8129}{{\tt arXiv:1412.8129 [hep-th]}}.

\bibitem{Chacaltana:2015bna}
O.~Chacaltana, J.~Distler, and A.~Trimm, ``{Tinkertoys for the Twisted $E_6$
  Theory},'' \href{http://dx.doi.org/10.1007/JHEP04(2015)173}{{\em JHEP} {\bf
  04} (2015)  173},
\href{http://arxiv.org/abs/1501.00357}{{\tt arXiv:1501.00357 [hep-th]}}.

\bibitem{Argyres:2007tq}
P.~C. Argyres and J.~R. Wittig, ``{Infinite coupling duals of N=2 gauge
  theories and new rank 1 superconformal field theories},''
  \href{http://dx.doi.org/10.1088/1126-6708/2008/01/074}{{\em JHEP} {\bf 01}
  (2008)  074},
\href{http://arxiv.org/abs/0712.2028}{{\tt arXiv:0712.2028 [hep-th]}}.

\bibitem{Argyres:2010py}
P.~C. Argyres and J.~Wittig, ``{Mass deformations of four-dimensional, rank 1,
  N=2 superconformal field theories},''
  \href{http://dx.doi.org/10.1088/1742-6596/462/1/012001}{{\em J. Phys. Conf.
  Ser.} {\bf 462} (2013) no.~1, 012001},
\href{http://arxiv.org/abs/1007.5026}{{\tt arXiv:1007.5026 [hep-th]}}.

\bibitem{Xie:2015rpa}
D.~Xie and S.-T. Yau, ``{4d N=2 SCFT and singularity theory Part I:
  Classification},''
\href{http://arxiv.org/abs/1510.01324}{{\tt arXiv:1510.01324 [hep-th]}}.

\bibitem{Aharony:2007dj}
O.~Aharony and Y.~Tachikawa, ``{A Holographic computation of the central
  charges of d=4, N=2 SCFTs},''
  \href{http://dx.doi.org/10.1088/1126-6708/2008/01/037}{{\em JHEP} {\bf 01}
  (2008)  037},
\href{http://arxiv.org/abs/0711.4532}{{\tt arXiv:0711.4532 [hep-th]}}.

\bibitem{Wang:2016yha}
Y.~Wang, D.~Xie, S.~S.~T. Yau, and S.-T. Yau, ``{4d N=2 SCFT from Complete
  Intersection Singularity},''
\href{http://arxiv.org/abs/1606.06306}{{\tt arXiv:1606.06306 [hep-th]}}.

\bibitem{Cecotti:2015hca}
S.~Cecotti and M.~Del~Zotto, ``{Higher S-dualities and Shephard-Todd groups},''
  \href{http://dx.doi.org/10.1007/JHEP09(2015)035}{{\em JHEP} {\bf 09} (2015)
  035},
\href{http://arxiv.org/abs/1507.01799}{{\tt arXiv:1507.01799 [hep-th]}}.

\bibitem{Caorsi:2016ebt}
M.~Caorsi and S.~Cecotti, ``{Homological S-Duality in 4d N=2 QFTs},''
\href{http://arxiv.org/abs/1612.08065}{{\tt arXiv:1612.08065 [hep-th]}}.

\bibitem{Shapere:1999xr}
A.~D. Shapere and C.~Vafa, ``{BPS structure of Argyres-Douglas superconformal
  theories},''
\href{http://arxiv.org/abs/hep-th/9910182}{{\tt arXiv:hep-th/9910182
  [hep-th]}}.

\bibitem{Chen:2016bzh}
B.~Chen, D.~Xie, S.-T. Yau, S.~S.~T. Yau, and H.~Zuo, ``{4d N=2 SCFT and
  singularity theory Part II: Complete intersection},''
\href{http://arxiv.org/abs/1604.07843}{{\tt arXiv:1604.07843 [hep-th]}}.

\end{thebibliography}\endgroup

\end{document}